\documentclass[lettersize,journal,10pt]{IEEEtran}
\usepackage{balance}
\usepackage[table]{xcolor}
\usepackage{xcolor}
\usepackage{soul}
\usepackage{mathtools}
\usepackage{epsfig}
\usepackage{color}
\usepackage{amsmath,amsthm,amssymb}
\usepackage{tikz}
\usepackage{breqn}
\usepackage{bbm}
\usepackage{gensymb}
\usepackage{url}

\usepackage{booktabs}

\usepackage{array}
\newcolumntype{M}[1]{>{\centering\arraybackslash}m{#1}}

\makeatletter
\renewcommand*\env@matrix[1][*\c@MaxMatrixCols c]{%
  \hskip -\arraycolsep
  \let\@ifnextchar\new@ifnextchar
  \array{#1}}
\makeatother

\usepackage{graphicx}
\usepackage{epstopdf}
\usepackage{algorithm}
\usepackage{algorithmic}
\usepackage{eucal}
\usepackage{enumerate}
\usepackage{cite}
\usepackage{amsfonts}
\usepackage{makecell}
\usepackage{lettrine}
\usepackage{caption}
\usepackage{amssymb}
\usepackage{blindtext}
\usepackage{multirow}
\usepackage{lipsum}
\usepackage[extra]{tipa}
\usepackage{comment}
\usepackage{colortbl}

\usepackage[font=small]{caption}

\def\BibTeX{{\rm B\kern-.05em{\sc i\kern-.025em b}\kern-.08em T\kern-.1667em\lower.7ex\hbox{E}\kern-.125emX}}
\usepackage[font=small,skip=5pt]{caption}
\newcolumntype{P}[1]{>{\centering\arraybackslash}p{#1}}

\usepackage{url}

\usepackage{breakurl}
\usepackage[breaklinks,hidelinks]{hyperref}

\begin{document}
%

\title{Sensor Allocation and Online-Learning-based Path Planning for Maritime Situational Awareness Enhancement: A Multi-Agent Approach}

\author{\IEEEauthorblockN{Bach Long Nguyen, Anh-Dzung Doan, Tat-Jun Chin, Christophe Guettier, Surabhi Gupta, Estelle Parra, Ian~Reid,~and~Markus~Wagner}\\
\thanks{Bach Long Nguyen and Markus Wagner are with the Department of Data Science and AI, Monash Univeristy, Clayton VIC 3800, Australia (email: long.nguyen1@monash.edu, markus.wagner@monash.edu).}
\thanks{Anh-Dzung Doan, Tat-Jun Chin and Ian Reid are with Australian Institute for Machine Learning, The University of Adelaide, Adelaide SA 5000, Australia (email: dzung.doan@adelaide.edu.au, tat-jun.chin@adelaide.edu.au, ian.reid@adelaide.edu.au).}
\thanks{Christophe Guettier and Estelle Parra are with Safran Electronics and Defense, Massy 91300, France (email: christophe.guettier@safrangroup.com, estelle.parra@safrangroup.com). }
\thanks{Surabhi Gupta is with Safran Electronics and Defense Australasia, Botany NSW 2019, Australia (email: surabhi.gupta@safrangroup.com). }
}

\markboth{IEEE Transactions on Intelligent Transportation Systems,~Vol.~XX, No.~XX,~2023}%
{Shell \MakeLowercase{\textit{et al.}}: A Sample Article Using IEEEtran.cls for IEEE Journals}


{}
\maketitle

\begin{abstract}
Countries with access to large bodies of water often aim to protect their maritime transport by employing maritime surveillance systems. However, the number of available sensors (e.g., cameras) is typically small compared to the to-be-monitored targets, and their Field of View (FOV) and range are often limited. 
This makes improving the situational awareness of maritime transports challenging. To this end, we propose a method that not only distributes multiple sensors but also plans paths for them to observe multiple targets, while minimizing the time needed to achieve situational awareness. In particular, we provide a formulation of this sensor allocation and path planning problem which considers the partial awareness of the targets' state, as well as the unawareness of the targets' trajectories. To solve the problem we present two algorithms: \emph{1)} a greedy algorithm for assigning sensors to targets, and \emph{2)} a distributed multi-agent path planning algorithm based on regret-matching learning. Because a quick convergence is a requirement for algorithms developed for high mobility environments, we employ a forgetting factor to quickly converge to correlated equilibrium solutions. Experimental results show that our combined approach achieves situational awareness more quickly than related work.
\end{abstract}


\IEEEpeerreviewmaketitle

\section{Introduction}


In Australia, 99\% of all exports rely on maritime transport; in recognition of this, the Departure of Home Affairs has recently commenced the Future Maritime Surveillance Capability project~\cite{Affair2023}. One of {}{those project aims} is to reduce current and emerging threats to Australian maritime transport. According to~\cite{Force2021}, threats include unauthorized maritime arrivals, maritime terrorism, and piracy, robbery or violence~at~sea.

In addition to governmental institutions, 
researchers have also increasingly been paying attention to the ocean~\cite{Zhou2021,Lin2022}. In particular, achieving maritime situational awareness is 
an important task for researchers when they are designing maritime surveillance systems. Such systems need to deal with complex and time-varying environments. Specifically, targets\footnote{We use the term ``target'' to identify any vessel that is of interest. Our work is strictly defensive in nature.}, such as fishing boats or sailing boats, can frequently change their directions, and their long-term trajectories are typically unknown. In addition, the number of available sensors, e.g., cameras, may be quite low (compared to the number of vessels present in commercial straits or harbours) while detecting range and FOV of cameras are limited. As a result, resource allocation and sensor path-planning in real time are essential to address these issues~\cite{Guerra2020,Lu2022,Ozturk2022}.  

In this paper, we consider scenarios where targets can initially be outside the coverage range of some sensors. Moreover, our sensors are cameras, radars and automatic identification systems (AIS\footnote{AIS: \url{https://en.wikipedia.org/wiki/Automatic_identification_system}}). 
However, unlike~\cite{Bloisi2017} and~\cite{Cormack2020}
, we place these cameras onboard of boats, e.g. guard boats, or unmanned surface vehicles (USVs). Thus, the sensors can move forward towards their targets in order to capture images and videos of these targets. Here, we assume that both the sensors and their targets move across the surface of a sea.  

\begin{figure}[!tbp]
        \centering
        \includegraphics[width=1\columnwidth]{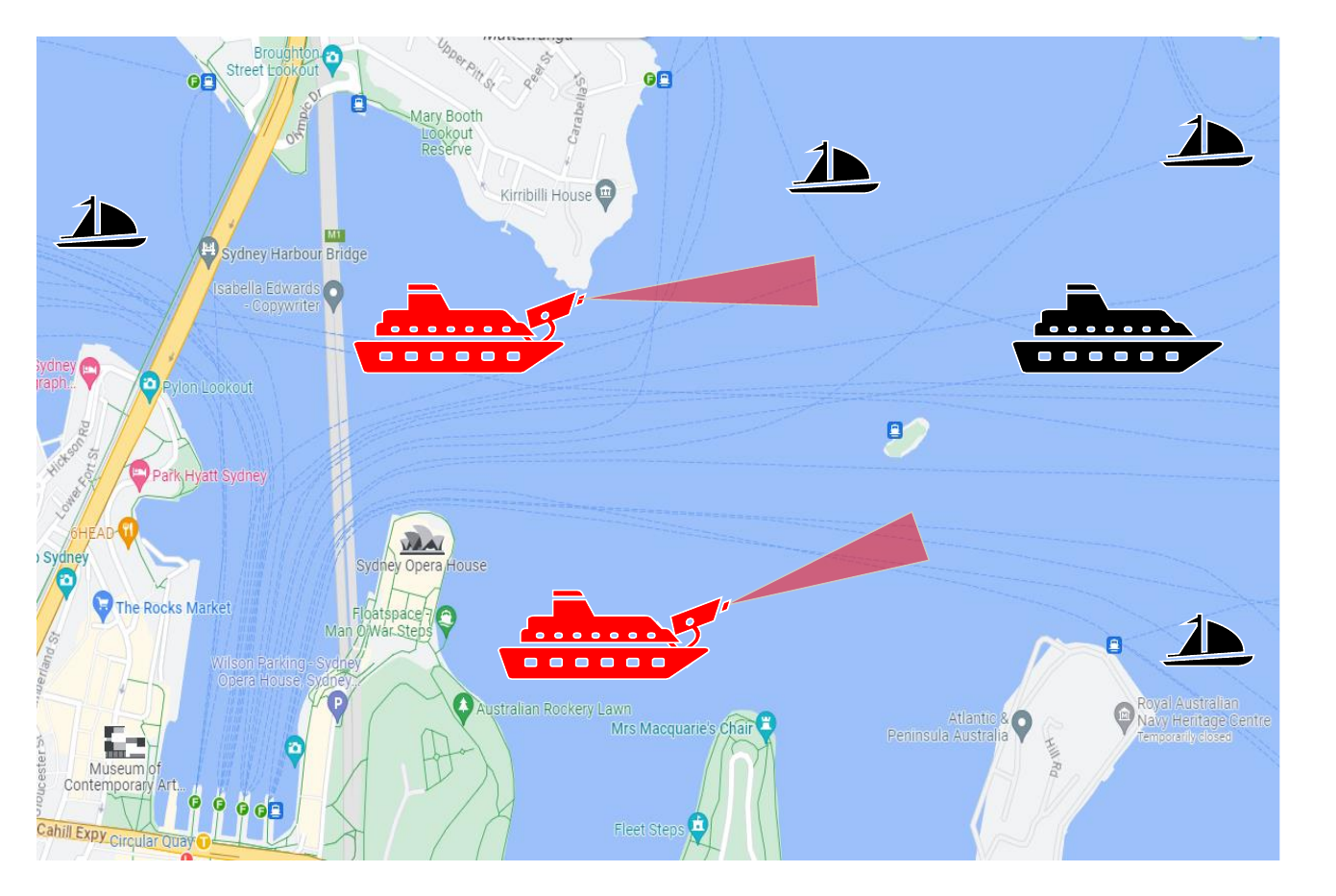}
        \captionsetup{justification=centering}
        \caption{An illustration of Sydney Harbor~\cite{Google2023}
        . In this scenario, \textit{1)} the two red boats are guard boats equipped with camera whose horizontal FOV is $3 \degree$, and \textit{2)} black vessels are targets which the two guard boats attempt to observe. Vessels and cameras are not drawn to scale.}
        \label{fig:scenario}
\end{figure}

We define by $\mathcal{C}$ and $\mathcal{V}$ a set of available cameras used and a set of targets, respectively. Thus, $|\mathcal{C}|$ is the number of cameras while $|\mathcal{V}|$ is the number of targets. Each target $v \in \mathcal{V}$ moves around in the area that is to be monitored, e.g., Sydney Harbor in Australia (see Fig.~\ref{fig:scenario}), and the target may initially not reside within the sensing range of any camera $c \in \mathcal{C}$. Here, the cameras are not only partially aware of targets' states\footnote{Partial awareness means that the cameras know that targets exist, but they do not know exactly where the targets are.}, but they are also unaware of target trajectories. Thus, we need to estimate the targets' locations over a time horizon. First, we assume that the motion of $|\mathcal{V}|$ targets is linear without any changes in moving speeds or directions. This assumption is reasonable because ships typically travel along sea lanes in harbors according to maritime vessel traffic~\cite{Imo2023}. Then, we rely on this assumption and our measurements that are collected by a stationary radar $r$ and the AIS 
to estimate future positions of targets.

In practice, radars and 
cameras can determine the range and bearing on their own 
, while AIS broadcast information about location and speed of vessels~\cite{Cormack2020}. Like~\cite{Takemura2007,Ragi2013,Bar2009}, we assume that the radar, AIS and cameras provide us with observations of the targets' location. Then, we estimate the targets' location using Kalman filters and probabilistic data association. Cameras in the set $\mathcal{C}$ will use the estimate to independently plan a path to move forward until the targets are inside their FOV. Here, they generate their own paths by picking a moving angle and a speed from a finite set of angles and a finite set of speed levels, respectively. 
In addition, we have to allocate the cameras to the targets and plan the cameras' trajectories. Here, we are partially aware of target state through measurements from the radar and AIS while the target trajectories are unknown. It makes both allocating limited resources and planning cameras' trajectories challenging. 






\subsection{Background}

Due to uncertainties in high mobility environments like maritime situations~\cite{Kisialiou2019,Capobianco2021,Thombre2022,Eva2022}, a number of research projects focus on maximizing sensors' coverage area or on planning paths for sensors to accurately sense targets. Specifically, \cite{Yoon2021} attempts to maximize sensor coverage in wireless sensor networks (WSN) using a local search-based algorithm. In addition, the authors mathematically determine the upper and lower bounds of coverage deployment. Compared to~\cite{Yoon2021}, who minimize the 
WSN cost,~\cite{Alia2017} uses a meta-heuristic algorithm to optimize the number of sensors and their locations. In contrast to these works, \cite{Zhang2017} considers uncertainties of response time and demand in the problem of locating emergency service facilities. Relying on an uncertainty theory, the authors model the maximal location problem under the uncertainties; then, they transform the model into an equivalent deterministic problem to solve it. However, all these works require knowledge of the target trajectories.

{}{To track targets moving with an unknown trajectory,~\cite{Wang2022_1,Wang2022_2,Wang2021,Jain2018,Shi2021,Cooper2018,Lu2021,Shi2021-2,Dogancay2012,Li2023} estimate targets' position and velocity, and then determine 
trajectory for USVs, autonomous underwater vehicles (AUVs) or UAVs. For example,~\cite{Wang2022_1,Wang2022_2,Wang2021} propose algorithms to track the trajectory of a single USV under unknown system dynamics.}  Based on range measurement,~\cite{Jain2018} predicts a target's future position based on its past trajectory and estimated velocity. Then, a reference path that closely follows the target's predicted path is generated. A controller uses feedback from the AUV's sensors to follow the reference path while maintaining a safe distance from the target. 
Similarly, \cite{Shi2021} and~\cite{Shi2021-2} employ data collection, target analysis, resource optimization or route optimization, while their goal is to improve the accuracy and effectiveness in tracking multiple targets. Interestingly, the work of~\cite{Shi2021} is effective only when sensors are stationary and when targets always stay inside the sensors' sensing range. 
As another limitation,~\cite{Shi2021-2} demonstrate a resource management scheme in the scenario where only one mobile sensor is employed. 

To plan paths for multiple mobile sensors or UAVs, \cite{Ragi2013,Floriano2019,Rizk2018,Hoa2020} propose algorithms based on the partially observable Markov decision process (POMDP) framework~\cite{Chades2021}. In~\cite{Ragi2013}, UAVs are guided to track multiple ground targets by selecting the best velocity and heading angle for the UAVs. That work also takes the effects of wind into account when controlling UAVs. Instead of designing a path-planning algorithm,~\cite{Hoa2020} focuses on determining the positions of UAVs in the future with the target which is to search and track multiple mobile objects. Both~\cite{Ragi2013} and~\cite{Hoa2020} do not resolve the issue where the targets do not traverse within the coverage range of the sensors initially. Unlike~\cite{Ragi2013,Hoa2020}, the work of~\cite{Li2022} introduces an algorithm which plans for multiple sensors to follow multiple targets leaving the sensors’ observation range.

One significant issue with the algorithms in~\cite{Ragi2013,Hoa2020,Li2022} is that they are only efficient (or demonstrated) in scenarios where the number of sensors and targets is limited to three and five, respectively. Additionally, these algorithms may not converge to an optimal solution. To tackle this issue, \cite{DDNguyen2017} and~\cite{CFan2020} have utilized regret-matching learning, a form of online learning, to create multi-agent system algorithms whose convergence is guaranteed. {}{According to~\cite{Hart2000}, regret-matching learning is a strategy designed for agents to learn how to make decisions over time. Specifically, each agent will play an action with respect to its probability, which is proportional to the gain of that action, and the action gain is measured by regrets.} The effectiveness of regret-matching learning-based algorithms is shown in various applications such as seasonal forecasting and matching markets without incentives~\cite{Xu2020,Flore2021,Genevieve2021,Bistritz2020}. Additionally, the regret-matching learning-based algorithms in~\cite{Daskalakis2022} and~\cite{Ioannis2022}, have been shown to converge to correlated-equilibrium solutions more quickly than reinforcement-learning-based algorithms. {}{However, these existing works~\cite{DDNguyen2017,CFan2020,Xu2020,Flore2021,Genevieve2021,Bistritz2020,Daskalakis2022,Ioannis2022} have not considered any problems related to allocation and planning.}

\subsection{Contribution}

In this paper, we propose a multi-sensor allocation and planning approach to observe multiple targets with unknown trajectories, while minimizing the time required to achieve situational awareness. 
Unlike~\cite{Ragi2013,Hoa2020,Li2022}, we consider that the number of available mobile cameras can be large, e.g., 30 cameras, and that it is lower than the number of targets. The targets are (at times) outside the limited FOV of cameras. Due to the partial awareness of target state, as well as the unawareness of target trajectory in the next time, planning multiple cameras' paths to observe all multiple targets during a short time in this scenario is a non-trivial problem. To solve this problem, we will assign cameras to targets; these cameras then make their own plan to move forward to the targets using regret-matching learning. We summarize the contributions of this paper in the following.

\begin{enumerate}
\item Unlike~\cite{Ragi2013}, we formulate a problem to not only assign cameras to targets but to also generate a trajectory for each camera over a time horizon for the given maritime situation. In particular, the partial awareness of target state, and the unawareness of target trajectory are taken into account in our problem formulation.

\item Due to the complexity and the large size of our formulated problem, i.e., a large number of sensors  and targets, or a long time horizon, we propose two algorithms: 
a sensor allocation algorithm and a distributed multi-agent path planning algorithm, to deal with the problem. Using the sensor-target distance and the duration that a target has not been observed, the former allocates cameras to targets to shorten the duration of observing all the targets. Subsequently, the latter employs 
regret-matching learning so that the sensors create their trajectory as individual agents in a multi-agent system. In our online-learning process, we use a forgetting factor to make the proposed distributed algorithm converge quickly to a correlated equilibrium,  
similar to our previous work~\cite{Long2022}.

\item We evaluate the performance using computational experiments that extend scenarios from the literature by several orders of magnitude. The results demonstrate the efficiency of our proposed approach to achieve situational awareness, especially in scenarios involving many cameras and targets. 


\end{enumerate}

The remainder of the paper is organized as follows. Section~\ref{sec:system model} presents the system model, including a dynamic model and a measurement model, while Section~\ref{sec:problem formulation} describes the problem formulation for sensor allocation and path planning over a time horizon. Then, we describe in Section~\ref{sec:proposed scheme} our multi-sensor allocation algorithm and our distributed multi-agent path planning algorithm. Section~\ref{sec:experiment results} reports on our experiments to illustrate the efficiency of the proposed approach. Finally, we summarize the paper in Section~\ref{sec:conclusion}.

\section{System Model} \label{sec:system model}

To formulate the problem, we need to model the mobility of targets, as well as the observations (measurements) of radars, AIS and cameras. The models of target motion and sensor measurement are described in the following subsections. 


\subsection{Dynamic Model}

Like~\cite{Cormack2020}, we assume that each target in our maritime scenario follows a nearly-constant velocity (NCV) model. The NCV model is given by:
\begin{equation}
    \setlength\abovedisplayskip{0pt}
    \setlength\belowdisplayskip{0pt}
    \begin{aligned}
        \bold{x}^v_{t+1} &= \bold{F}_t \cdot \bold{x}^v_t + \bold{w}^v_t 
    \end{aligned}
    \label{eq:target_motion_model}
\end{equation}
with
\begin{equation}
    \setlength\abovedisplayskip{0pt}
    \setlength\belowdisplayskip{0pt}
\begin{aligned}
        \bold{x}^v_t &=(x^v_t \quad \Dot{x}^v_t \quad y^v_t 
        \quad \Dot{y}^v_t)^T,\\
        \bold{F}_t&= \begin{pmatrix}
            1&\Delta t&0&0\\
            0&1&0&0\\
            0&0&1&\Delta t\\
            0&0&0&1
        \end{pmatrix},
    \end{aligned}
\end{equation}
where $x^v_t$ and $y^v_t$ are target $v$'s position while $\Dot{x}^v_t$ and $\Dot{y}^v_t$ are target $v$'s velocity at time step $t$ in a 2-D coordinate system. $\bold{x}^T$ is the transposition of vector $\bold{x}$. In addition, $\Delta{t}$ is the duration of each time step {}{while $\bold{F}_t$ is a motion model for all targets}. $\bold{w}^v_t \sim N(0,\bold{Q}^v_t)$ is zero-mean white Gaussian process noise with the following covariance $\bold{Q}^v_t$.
\begin{equation}
    \setlength\abovedisplayskip{0pt}
    \setlength\belowdisplayskip{0pt}
    \begin{aligned}
        \bold{Q}^v_t&= {\sigma_v}^2\begin{bmatrix}
            \frac{(\Delta t)^3}{3}&\frac{(\Delta t)^2}{2}&0&0\\
            \frac{(\Delta t)^2}{2}&\frac{(\Delta t)^3}{3}&0&0\\
            0&0&\frac{(\Delta t)^3}{3}&\frac{(\Delta t)^2}{2}\\
            0&0&\frac{(\Delta t)^2}{2}&\frac{(\Delta t)^3}{3}
        \end{bmatrix}.
    \end{aligned}
\end{equation}

\subsection{Measurement Model}

Like~\cite{Ragi2013,Li2022}, we model the observations of the radar $r$, AIS and each camera $c \in \mathcal{C}$, respectively, in the following.
\begin{equation}
    \setlength\abovedisplayskip{0pt}
    \setlength\belowdisplayskip{0pt}
    \begin{aligned}
        &\bold{z}^{rv}_t \quad &= \quad  \bold{H}^r_t \cdot \bold{x}^{v}_t + \bold{v}^{rv}_t\\
        &\bold{z}^{\text{AIS}v}_t \quad &= \quad  \bold{H}^{\text{AIS}}_t \cdot \bold{x}^{v}_t + \bold{v}^{\text{AIS}}_t\\
        &\bold{z}^{cv}_t \quad &= \quad  \bold{H}^c_t \cdot \bold{x}^{v}_t + \bold{v}^{cv}_t
    \end{aligned}
    \label{eq:observations}
\end{equation}
with
\begin{equation}
    \setlength\abovedisplayskip{0pt}
    \setlength\belowdisplayskip{0pt}
    \begin{aligned}
        \bold{H}^r_t = \bold{H}^{\text{AIS}}_t = \bold{H}^c_t &= \begin{pmatrix}
            1&0&0&0\\
            0&0&1&0
        \end{pmatrix},
    \end{aligned}
\end{equation}
where {}{$\bold{H}^r_t$, $\bold{H}^{\text{AIS}}_t$ and $\bold{H}^c_t$ are the observation models of radar $r$, AIS and camera $c$, respectively. Moreover,} $\bold{v}^{rv}_t \sim N(0,\bold{R}^{rv}_t)$, $\bold{v}^{\text{AIS}}_t \sim N(0,\bold{R}^{\text{AIS}}_t)$ and $\bold{v}^{cv}_t \sim N(0,\bold{R}^{cv}_t)$ are additive noise terms. Then, $\bold{R}^{rv}_t$, $\bold{R}^{\text{AIS}}_t$ and $\bold{R}^{cv}_t$ are determined as follows:
\begin{equation}
    \setlength\abovedisplayskip{0pt}
    \setlength\belowdisplayskip{0pt}
    \begin{aligned}
        &\bold{R}^{rv}_t &=\quad {{\sigma}^{rv}_t}^2
        \begin{pmatrix}
            1 & 0 \\
            0 & 1
        \end{pmatrix}\\
        &\bold{R}^{\text{AIS}}_t &=\quad {{\sigma}^{\text{AIS}}}^2
        \begin{pmatrix}
            1 & 0 \\
            0 & 1
        \end{pmatrix}\\
        &\bold{R}^{cv}_t &=\quad {{\sigma}^{cv}_t}^2
        \begin{pmatrix}
            1 & 0 \\
            0 & 1
        \end{pmatrix}. 
    \end{aligned}
    \label{eq:measurement_model}
\end{equation}
In a maritime environment, only uncertainties in the observations of radars and cameras depend on the distance between these observers and their targets. Thus, in Eq.~\eqref{eq:measurement_model},  ${\sigma^{rv}_t}$ and ${\sigma^{cv}_t}$ are positive proportional to the distance from radar $r$ or camera $c$ to target $v$ at time step $t$ while ${\sigma^{\text{AIS}}}$ is assumed to be constant (because it is GPS-based). 

\section{Problem Formulation} \label{sec:problem formulation}

To assign targets to cameras and to plan trajectories for these cameras (possibly over a long time horizon), we have to estimate the targets' positions using the observations collected by radar $r$, AIS and cameras. 
Furthermore, in this paper, we consider that the overall maritime situational awareness is achieved (albeit temporarily) once all targets are observed at least once\footnote{This is merely a practical consideration. As we will demonstrate, our approach can handle unbounded scenarios.}. 
Here, we define that a target is observed by a camera if it resides within the camera' detection range and FOV, and the mean squared error (MSE) between its estimated value and ground truth is equal to or less than a pre-given threshold. 

{In this paper, a target $v$ can be observed by radar $r$, AIS and camera $c$; thus, by building on~\cite{Ragi2013}, the total MSE of $|\mathcal{V}|$ targets is calculated over a time horizon $H$ as follows:}
\begin{equation}
        \setlength\abovedisplayskip{0pt}
        \setlength\belowdisplayskip{0pt}
    \begin{aligned}
    J_H &\approx \sum_{t=\Tilde{t}+0}^{\Tilde{t}+H-1} \sum_{v\in \mathcal{V}} \text{Tr} \left(\Tilde{\bold{P}}^v_{t+1}\right)\\
    & = \sum_{t=\Tilde{t}+0}^{\Tilde{t}+H-1} \sum_{v\in \mathcal{V}} \text{Tr} \left( \frac{1}{\frac{1}{\Tilde{\bold{P}}^{rv}_{t+1}}+\frac{1}{\Tilde{\bold{P}}^{\text{AIS}v}_{t+1}}+\frac{1}{\Tilde{\bold{P}}^{cv}_{t+1}}} \right),
    \end{aligned}
\label{eq:new-objective_function}
\end{equation}
\begin{figure}[!tbp]
        \centering \includegraphics[width=1\columnwidth]{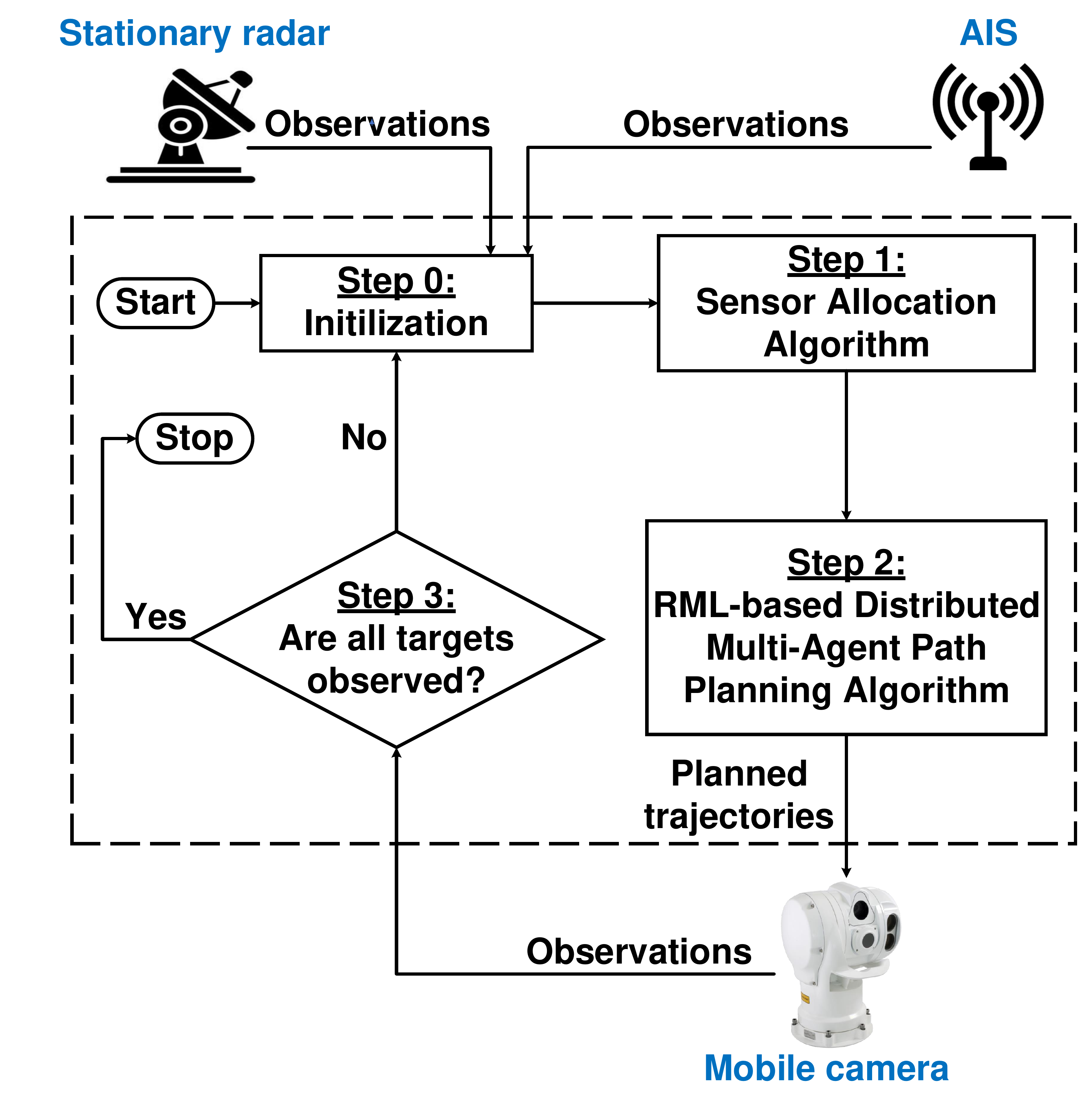}
        \captionsetup{justification=centering}
        \caption{Flowchart of the proposed Sensor Allocation and Path Planning approach.}
        \label{fig:flowchart}
\end{figure}
where $t=\Tilde{t}+0$ is the current time step; and $\text{Tr} \left(\Tilde{\bold{P}}^v_{t+1}\right)$ is the trace of matrix $\Tilde{\bold{P}}^v_{t+1}$. {}{Here, $\Tilde{\bold{P}}^v_{t+1}$ is the covariance update of target $v$ at time step $t+1$.} To compute $\Tilde{\bold{P}}^{rv}_{t+1}$, $\Tilde{\bold{P}}^{\text{AIS}v}_{t+1}$ and $\Tilde{\bold{P}}^{cv}_{t+1}$, we have
\begin{equation}
        \setlength\abovedisplayskip{0pt}
        \setlength\belowdisplayskip{0pt}
\begin{aligned}
    &\Tilde{\bold{P}}^{rv}_{t+1}=
    \begin{cases}
        \frac{1}{\frac{1}{\Tilde{\bold{P}}^{rv}_{t+1|t}}+\bold{S}^{rv}_{t+1}} & \text{if $v$ is observed by $r$},\\
        \Tilde{\text{P}}^{rv}_{t+1|t} & \text{otherwise} 
    \end{cases}\\
    &\Tilde{\bold{P}}^{\text{AIS}v}_{t+1}=
    \begin{cases}
        \frac{1}{\frac{1}{\Tilde{\bold{P}}^{\text{AIS}v}_{t+1|t}}+\bold{S}^{\text{AIS}v}_{t+1}} & \text{if $v$ is observed by AIS},\\
        \Tilde{\bold{P}}^{\text{AIS}v}_{t+1|t} & \text{otherwise} 
    \end{cases}\\
    &\Tilde{\bold{P}}^{cv}_{t+1}=
    \begin{cases}
        \frac{1}{\frac{1}{\Tilde{\bold{P}}^{cv}_{t+1|t}}+\bold{S}^{cv}_{t+1}} & \text{if $v$ is observed by $c$},\\
        \Tilde{\bold{P}}^{cv}_{t+1|t} & \text{otherwise} 
    \end{cases},
\end{aligned}
\label{eq:covariance}
\end{equation}
where $\Tilde{\bold{P}}^{rv}_{t+1|t}=\bold{F}_t \cdot \Tilde{\bold{P}}^{rv}_{t} \cdot \bold{F}^T_t + \bold{Q}^v_t$, $\Tilde{\bold{P}}^{\text{AIS}v}_{t+1|t}=\bold{F}_t \cdot \Tilde{\bold{P}}^{\text{AIS}v}_{t} \cdot \bold{F}^T_t + \bold{Q}^v_t$ and $\Tilde{\bold{P}}^{cv}_{t+1|t}=\bold{F}_t \cdot \Tilde{\bold{P}}^{cv}_{t} \cdot \bold{F}^T_t + \bold{Q}^v_t$. Additionally, {}{$\bold{S}^{rv}_{t+1}$, $\bold{S}^{\text{AIS}v}_{t+1}$ and $\bold{S}^{cv}_{t+1}$ are the measurement prediction covariance of radar $r$, AIS and camera $c$ when they observe target $v$. Then, we have} $\bold{S}^{rv}_{t+1}={\bold{H}^r_{t+1}}^T \cdot \bold{R}^{rv}_{t+1} \cdot \bold{H}^r_{t+1}$, $\bold{S}^{\text{AIS}v}_{t+1}={\bold{H}^{\text{AIS}}_{t+1}}^T \cdot \bold{R}^{\text{AIS}}_{t+1} \cdot \bold{H}^{\text{AIS}}_{t+1}$ and $\bold{S}^{cv}_{t+1}={\bold{H}^c_{t+1}}^T \cdot \bold{R}^{cv}_{t+1} \cdot \bold{H}^c_{t+1}$. We assume that target $v$ will be observed by radar $r$, AIS and camera $c$ at time step $t+1$ if target $v$'s nominal mean positions defined by $\Tilde{\xi}^{rv}_{t+1}$, $\Tilde{\xi}^{\text{AIS}v}_{t+1}$ and $\Tilde{\xi}^{cv}_{t+1}$ are in the coverage area of radar $r$ and AIS, and the FOV of camera $c$, respectively. Similar to~\cite{Ragi2013}, we calculate $\Tilde{\xi}^{rv}_{t+1}$, $\Tilde{\xi}^{\text{AIS}v}_{t+1}$ and $\Tilde{\xi}^{cv}_{t+1}$, respectively, as follows: 
\begin{equation}
    \setlength\abovedisplayskip{0pt}
    \setlength\belowdisplayskip{0pt}
    \begin{aligned}
    \Tilde{\xi}^{rv}_{t+1}\quad &=\quad \bold{F}_t \cdot \Tilde{\xi}^{rv}_{t},\\
    \Tilde{\xi}^{\text{AIS}v}_{t+1}\quad &=\quad \bold{F}_t \cdot \Tilde{\xi}^{\text{AIS}v}_{t},\\
    \Tilde{\xi}^{cv}_{t+1}\quad &=\quad \bold{F}_t \cdot \Tilde{\xi}^{cv}_{t}.
    \end{aligned}
    \label{eq:mean}
\end{equation} When $t=\Tilde{t}+0$, ($\Tilde{\xi}^{rv}_{t}$, $\Tilde{\bold{P}}^{rv}_{t}$), ($\Tilde{\xi}^{\text{AIS}v}_{t}$,$\Tilde{\bold{P}}^{\text{AIS}v}_{t}$) and ($\Tilde{\xi}^{cv}_{t}$,$\Tilde{\bold{P}}^{cv}_{t}$) are the state update and covariance update of target $v$ tracked by radar $r$, AIS and camera $c$, respectively. In addition, we calculate ($\Tilde{\xi}^{rv}_{t}$, $\Tilde{\bold{P}}^{rv}_{t}$), ($\Tilde{\xi}^{\text{AIS}v}_{t}$,$\Tilde{\bold{P}}^{\text{AIS}v}_{t}$) and ($\Tilde{\xi}^{cv}_{t}$,$\Tilde{\bold{P}}^{cv}_{t}$) using Kalman filter and probabilistic data association. 

{By using Eqs.~\eqref{eq:new-objective_function},~\eqref{eq:covariance} and~\eqref{eq:mean}, we formulate the optimization problem of multi-sensor allocation and path planning over a time horizon $H$ in the following.}

\begin{subequations}
    \label{origin problem}
    \begin{alignat}{6}
        \min_{\Hat{x}^{cv}_t,s^c_t,\phi^c_t} \quad 
        \begin{split}
            &\sum_{t=\Tilde{t}+0}^{\Tilde{t}+H-1}\Bigg\{\sum_{v\in \mathcal{V}} \Big[ \Big(\text{Tr} \left(\doubletilde{{P}}^v_{t+1}\right) - \epsilon \Big) \\
            &+\alpha_1\sum_{c\in \mathcal{C}} \Hat{x}^{cv}_t(D_{\min}-\Tilde{d}^{cv}_{t+1})(\Tilde{d}^{cv}_{t+1}-D_{\max})\\
            &+\alpha_2 \sum_{c\in \mathcal{C}}(D_{\text{safe}}-\Tilde{d}^{cv}_{t+1})
            \Big]\\
            &+\alpha_3\sum_{c\in \mathcal{C}}\sum_{c'\in\mathcal{C},c'\neq c} (D_{\text{safe}}-d^{cc'}_{t+1})
            \Bigg\}
        \end{split} \label{objective function problem}
        \\
        \text{s.t.} \quad \begin{split}
            &s_{\min} \leq s^c_t \leq s_{\max} \label{con-1}\\
            &\quad \quad \forall c \in \mathcal{C}; t=\Tilde{t}+0,\dots,\Tilde{t}+H-1
            \end{split}\\
            \begin{split}
            & 0 \leq |\phi^c_{t}-\phi^c_{t-1}| \leq \Phi_{\max} \label{con-2}\\
            &\quad \quad \forall c \in \mathcal{C}; t=\Tilde{t}+0,\dots,\Tilde{t}+H-1
            \end{split}\\
            \begin{split}
            &\sum_{c \in \mathcal{C}} x^{cv}_t \leq 1\\
            &\quad \quad  \forall v \in \mathcal{V}; t=\Tilde{t}+0,\dots,\Tilde{t}+H-1 \label{con-3}
            \end{split}\\
            \begin{split}
            & \sum_{v \in \mathcal{V}} x^{cv}_t = 1\\
            &\quad \quad \forall c \in \mathcal{C}; t=\Tilde{t}+0,\dots,\Tilde{t}+H-1 \label{con-4}
            \end{split}
    \end{alignat}
\end{subequations}
with
\begin{equation}
\setlength\abovedisplayskip{0pt}
\setlength\belowdisplayskip{0pt}
\doubletilde{{P}}^v_{t+1}=\frac{1}{\frac{1}{\Tilde{\mathbf{P}}^{rv}_{t+1}}+\frac{1}{\Tilde{\mathbf{P}}^{\text{AIS}v}_{t+1}}+\sum_{c\in \mathcal{C}}\frac{\Hat{x}^{cv}_t}{\Tilde{\mathbf{P}}^{cv}_{t+1}}},
\end{equation}
where $\text{Tr} \left(\doubletilde{{P}}^v_{t+1}\right)$ is the trace of matrix $\doubletilde{{P}}^v_{t+1}$; and $\epsilon$ is the MSE threshold. We use $\Hat{x}^{cv}_t$ for assigning a target $v$ to a camera $c$. If $\Hat{x}^{cv}_t=1$, target $v$ is assigned to camera $c$. Otherwise, target $v$ is not assigned to camera $c$. Moreover, $s^c_t$ and $\phi^c_t$ are speed and moving angle of camera $c$, respectively, at time step $t$.
Different from~\cite{Ragi2013,Nguyen2019}, the objective function in~\eqref{objective function problem} includes the following four terms: \textit{1)} MSE, \textit{2)} the distance between the position of camera $c$ and the nominal mean position of target $v$ assigned to $c$, \textit{3)} the distance between the nominal mean position of target $v$ and the position of all cameras, and \textit{4)} the distance between a camera and another cameras at time step $t+1$. Here, the second term is employed to maintain that camera $c$ has to move forward until its target $v$ resides within its FOV if $c$ wants to observe $v$. Then, because cameras and targets move on the same surface (the surface of a sea), we use the third and fourth terms to guarantee a safety distance between cameras and targets. In addition, we have

\begin{subequations}
    \begin{equation}
            \setlength\abovedisplayskip{0pt}
        \setlength\belowdisplayskip{0pt}
        \text{Tr} \left(\doubletilde{{P}}^v_{t+1}\right) - \epsilon = 
        \begin{cases}
            0 & \text{if $\text{Tr} \left(\doubletilde{{P}}^v_{t+1}\right) \leq \epsilon$}, \\
            \text{Tr} \left(\doubletilde{{P}}^v_{t+1}\right) - \epsilon & \text{otherwise}
        \end{cases}
    \end{equation}
    \begin{equation}
            \setlength\abovedisplayskip{0pt}
        \setlength\belowdisplayskip{0pt}
        D_{\min}-\Tilde{d}^{cv}_{t+1} = 
        \begin{cases}
            1 & \text{if $\Tilde{d}^{cv}_{t+1} \geq D_{\min}$}, \\
            D_{\min}-\Tilde{d}^{cv}_{t+1} & \text{otherwise}
        \end{cases}
    \end{equation}
    \begin{equation}
            \setlength\abovedisplayskip{0pt}
        \setlength\belowdisplayskip{0pt}
        \Tilde{d}^{cv}_{t+1} - D_{\max} = 
        \begin{cases}
            1 & \text{if $\Tilde{d}^{cv}_{t+1} \leq D_{\max}$}, \\
            \Tilde{d}^{cv}_{t+1} - D_{\max} & \text{otherwise}
        \end{cases}
    \end{equation}
    \begin{equation}
            \setlength\abovedisplayskip{0pt}
        \setlength\belowdisplayskip{0pt}
        D_{\text{safe}} - \Tilde{d}^{cv}_{t+1} = 
        \begin{cases}
            0 & \text{if $\Tilde{d}^{cv}_{t+1} \geq D_{\text{safe}}$}, \\
            D_{\text{safe}} - \Tilde{d}^{cv}_{t+1} & \text{otherwise}
        \end{cases}
    \end{equation}
    \begin{equation}
            \setlength\abovedisplayskip{0pt}
        \setlength\belowdisplayskip{0pt}
        D_{\text{safe}} - d^{cc'}_{t+1}= 
        \begin{cases}
            0 & \text{if $d^{cc'}_{t+1} \geq D_{\text{safe}}$}, \\
            D_{\text{safe}} - d^{cc'}_{t+1} & \text{otherwise}
        \end{cases}
    \end{equation}
\end{subequations}
where the values of $D_{\min}$, $D_{\max}$ and $D_{\text{safe}}$ are pre-given; and $\alpha_1$, $\alpha_2$ and $\alpha_3$ are scaling factors $(\alpha_1,\alpha_2,\alpha_3 \in [0;1])$. Additionally, $\Tilde{d}^{cv}_{t+1}$ is the distance between the position of camera $c$ and the nominal mean position $\Tilde{\xi}^{cv}_{t+1}$ of target $v$ while $d^{cc'}_{t+1}$ is the distance between camera $c$ and camera $c'$ at time step $t+1$. Given $s^c_t$ and $\phi^c_t$, the position of camera $c$ at time step $t+1$ is determined using the following movement model, which is applicable to maritime situations~\cite{Han2020}.
\begin{equation}
        \setlength\abovedisplayskip{0pt}
        \setlength\belowdisplayskip{0pt}
        \begin{aligned}
            \begin{cases}
                x^c_{t+1} = x^c_t + s^c_t \cdot \text{cos}(\phi^c_t) \cdot 
                \Delta t\\
                y^c_{t+1} = y^c_t + s^c_t \cdot \text{sin}(\phi^c_t) \cdot \Delta t
            \end{cases},            
        \end{aligned}
        \label{eq:camera_motion_model}
\end{equation}
where $x^c_t$ and $y^c_t$ are positions of camera $c$ in a 2-D coordinate system (the sea surface) at time step $t$.
Conditions~\eqref{con-1} and~\eqref{con-2} describe the range for speed and moving
angle of each mobile camera $c$. Condition~\eqref{con-3} 
shows that one target must not be assigned to more than one camera at each time step because the number of cameras is inadequate. Moreover, condition~\eqref{con-4} defines 
that a camera only approaches a target until the target resides within 
{}{the camera FOV}. This constraint mimics human observer who focuses on one target at a time. 

\begin{algorithm}[t]
\scriptsize
    \caption{\textbf{The proposed Sensor Allocation and RML-based distributed Path Planning approach}}
    \begin{algorithmic}[1]
        \STATE \textbf{Input:} A camera set $\mathcal{C}$, a target set $\mathcal{V}$, the current position of $\mathcal{C}$ cameras, and  the information about $\mathcal{V}$ targets collected up to the current time step $t=\Tilde{t}+0$;
        \STATE \textbf{Output:} Camera-target allocation~$\Hat{x}^{cv}_t$, camera speed~$s^c_t$, and camera moving angle~$\phi^c_t$ ;
        \STATE \textbf{Step 0 (Initialization):}
        \STATE \quad Calculate nominal mean and covariance~($\Tilde{\xi}^{rv}_{t}$,$\Tilde{\bold{P}}^{rv}_{t}$), ($\Tilde{\xi}^{\text{AIS}v}_{t}$,$\Tilde{\bold{P}}^{\text{AIS}v}_{t}$) and \\ \quad \> ($\Tilde{\xi}^{cv}_{t}$,$\Tilde{\bold{P}}^{cv}_{t}$) at the current time step $t=\Tilde{t}+0$ using Kalman filter and \\ \quad \> probabilistic data association;
        \STATE \textbf{Step 1 (Multi-Sensor Allocation Algorithm):}
        \STATE \quad \textbf{for} $t=\Tilde{t}+0,\Tilde{t}+1,\dots,\Tilde{t}+H-1$
        \STATE \quad \quad \textbf{If} $t=\Tilde{t}+0$
        \STATE \quad \quad \quad Compute the allocation matrix $\mathbf{M}_t$ using Eqs.~\eqref{eq:beta} and~\eqref{eq:matrix_m};
        \STATE \quad \quad \quad Calculate $\Hat{x}^{cv}_t$ using the $3$-step process in section~\ref{sensor allocation};
        \STATE \quad \quad \textbf{else}
        \STATE \quad \quad \quad $\Hat{x}^{cv}_t=\Hat{x}^{cv}_{t-1}$;
        \STATE \quad \quad \textbf{end}
        \STATE \quad \textbf{end for}
        \STATE \textbf{Step 2 (RML-based Distributed Multi-Agent Path Planning Algorithm):}
        \STATE \quad \textbf{for} $t=\Tilde{t}+0,\Tilde{t}+1,\dots,\Tilde{t}+H-1$
        \STATE \quad \quad Calculate nominal mean and covariance~($\Tilde{\xi}^{rv}_{t+1}$,$\Tilde{\bold{P}}^{rv}_{t+1}$), ($\Tilde{\xi}^{\text{AIS}v}_{t+1}$,$\Tilde{\bold{P}}^{\text{AIS}v}_{t+1}$) and \\ \quad \quad \>  ($\Tilde{\xi}^{cv}_{t+1}$,$\Tilde{\bold{P}}^{cv}_{t+1}$) using Eqs.~\eqref{eq:covariance} and~\eqref{eq:mean};
        \STATE \quad \quad \textbf{Regret-matching procedure} run independently by each camera $c \in \mathcal{C}$:
        \STATE \quad \quad \textbf{\{} Initialize the probability distribution at $\tau=1$ of time step $t$:\\
        \quad \quad \quad \> $\pi^c_{t,1}(j) \leftarrow \frac{1}{|\mathcal{A}^c_{t,1}|}$ $\forall j\in \mathcal{A}^c_{t,1}$  
        \STATE \quad \quad \quad \textbf{for} $\tau=1,2,\dots,T$
        \STATE \quad \quad \quad \quad \textbf{Action selection:}       
        \STATE \quad \quad \quad \quad \quad Select action $a^c_{t,\tau}=j$ from its probability distribution $\pi_{t,\tau}^c(j).$
        \STATE \quad \quad \quad \quad \quad Update to all other cameras with the action that has been chosen.\\
        \STATE \quad \quad \quad \quad \textbf{Utility update:} 
        \STATE \quad \quad \quad \quad \quad Given the decisions made by the other cameras~$\overline{c}$, calculate the \\ \quad \quad \quad \quad \quad \> utility $u^c_{t,\tau}(j,a^{\overline{c}}_{t,\tau})$ using Eq.~\eqref{eq:utility}.
        \STATE \quad \quad \quad \quad \textbf{for} $k\in \mathcal{A}^c_{t,\tau}\setminus\{j\}$
        \STATE \quad \quad \quad \quad \quad \textbf{Expected utilities:} 
        \STATE \quad \quad \quad \quad \quad \quad Given the decisions made by the other cameras~$\overline{c}$, calculate the \\ \quad \quad \quad \quad \quad \quad \> expected utility $u^c_{t,\tau}(k,a^{\overline{c}}_{t,\tau})$ using Eq.~\eqref{eq:utility}.
        \STATE \quad \quad \quad \quad \quad \textbf{Regret update:} 
        \STATE \quad \quad \quad \quad \quad \quad Determine the cumulative regret $\Hat{D}^c_{t,\tau}(j,k)$ for not choosing action \\ \quad \quad \quad \quad \quad \quad \> $k$ using Eq.~\eqref{eq:probability}.
        \STATE \quad \quad \quad \quad \quad \textbf{Strategy update:}
        \STATE \quad \quad \quad \quad \quad \quad Compute the next action probability $\pi^c_{t,\tau+1}(k)$ using Eq.~\eqref{eq:updated_probability}.
        \STATE \quad \quad \quad \quad \quad \quad Compute the probability that camera $c$ plays the same action selected \\ \quad \quad \quad \quad \quad \quad \> in the current iteration using Eq.~\eqref{eq:rml}.
        \STATE \quad \quad \quad \quad \textbf{end for}
        \STATE \quad \quad \quad \textbf{end for}
        \STATE \quad \quad \quad Pick action $a^c_{t,T+1}=j$ according to the probability \\ \quad \quad \quad \> distribution $\pi^c_{t,T+1}(j)$.
        \STATE \quad \quad \quad $s^c_t=s^c_{t,T+1}$ and $\phi^c_t=\phi^c_{t,T+1}$. \textbf{\}}
        \STATE \quad \textbf{end for}
        \STATE \textbf{Step 3 (Stopping test):}
        \STATE \quad \textbf{If} all $|\mathcal{V}|$ targets are observed
        \STATE \quad \quad Stop \textbf{Algorithm~\ref{alg:RML}};
        \STATE \quad \textbf{else}
        \STATE \quad \quad Go to \textbf{Step 0};
        \STATE \quad \textbf{end}
    \end{algorithmic}
    \label{alg:RML}
\end{algorithm}












\section{Proposed Sensor Allocation and Path Planning Approach} \label{sec:proposed scheme}

It is hard to solve the problem~\eqref{origin problem} because of its complexity and its large search space, i.e., ~$\left[(|\mathcal{C}|\times|\mathcal{V}|)\times(|\mathcal{L}_s|\times|\mathcal{L}_m|)^{|\mathcal{C}|}\right]^H$ where $|\mathcal{L}_s|$ and $|\mathcal{L}_m|$ are the number of speed levels and heading angles of each camera. To this end, we design two separate algorithms: a sensor allocation algorithm and a distributed multi-agent path planning algorithm. Based on a greedy algorithm, the former distributes cameras to targets to reduce the duration of sensing all the targets. Moreover, in the latter, each camera is considered as an agent in a multi-agent system. These agents individually plan their trajectory using an online-learning method called regret-matching learning (RML). We summarize the proposed camera allocation and path planning approach in Fig.~\ref{fig:flowchart}. Furthermore, we present the pseudo-code of {}{the proposed Sensor Allocation and RML-based distributed Path Planning approach (SAPP)} in Alg.~\ref{alg:RML}.

\subsection{Multi-Sensor Allocation Scheme} \label{sensor allocation}

Due to the limited number of mobile cameras, these cameras have to be distributed to targets at the current time step $t=\Tilde{t}+0$ before planning their trajectory. Unlike~\cite{Zhijun2005,Huang2020,Han2020}, we introduce the following metric as a criteria of allocating a camera to a target at time step $t=\Tilde{t}+0$. 
\begin{equation}
        \setlength\abovedisplayskip{0pt}
        \setlength\belowdisplayskip{0pt}
        \beta^{cv}_t=\frac{\Delta\Tilde{\tau}^v_t}{d^{cv}_t}  \quad
        \substack{\forall c \in \mathcal{C}\\ \forall v \in \mathcal{V}},
        \label{eq:beta}
\end{equation}
where $\Delta\Tilde{\tau}^v_t$ describes how long target $v$ has not been observed by any cameras until time step $t$. If target $v$ is observed at a time step $t$, $\Delta\Tilde{\tau}^v_t$ will be reset to $0$. Otherwise, it increases by one time step. Moreover, $d^{cv}_t$ is the distance between camera $c$ and target $v$ at time step $t=\Tilde{t}+0$. Here, we consider the location of target $v$ at time step $t=\Tilde{t}+0$ as $\Tilde{\xi}^{cv}_t$, which is computed using Kalman filter and probabilistic data association. Consequently, a $|\mathcal{C}|$-by-$|\mathcal{V}|$ matrix $\bold{\text{M}}_t$  with $\beta^{cv}_t$ as an element is computed as follows:
\begin{equation}
        \setlength\abovedisplayskip{0pt}
        \setlength\belowdisplayskip{0pt}
    \bold{\text{M}}_t =  
    \begin{pmatrix}
        \beta^{11}_t & \hdots & \beta^{1|\mathcal{V}|}_t\\
        \vdots             & \ddots       &  \vdots \\                       
        \beta^{|\mathcal{C}|1}_t & \hdots & \beta^{|\mathcal{C}||\mathcal{V}|}_t 
    \end{pmatrix}.
    \label{eq:matrix_m}
\end{equation}
Using the matrix $\bold{\text{M}}_t$, the $3$-step process where we allocate a camera to a target is summarized in the following.
\begin{itemize}
    \item \underline{Step 1.1}: We will assign a camera $c$ to a target $v$ $(\Hat{x}^{cv}_t=1)$ if their $\beta^{cv}_t$ is the maximum element in $\bold{\text{M}}_t$.

    \item \underline{Step 1.2}: The other elements $\beta^{cv}_t$ relating to this camera $c$ and this target $v$ are removed from $\bold{\text{M}}_t$.

    \item \underline{Step 1.3}: If all $|\mathcal{C}|$ cameras are allocated to targets,
    we stop the process. Otherwise, we go back to Step 1.1. 
\end{itemize}
In the proposed sensor allocation scheme, the sensor allocation depends not only on the distance between observers and their targets but also on the duration for which the targets have not been looked at by any camera. Therefore, this proposed scheme is helpful for reducing the duration of observing all the targets at least once.

\subsection{RML-based Distributed Multi-Agent Path Planning Scheme}

After targets are assigned to cameras, we need to plan trajectories for the cameras to move towards their targets. This trajectory planning problem is reformulated as a multi-agent distributed problem. Here, we consider each mobile camera as an agent which is an independent decision maker. As soon as these agents perform an action, they will update the other agents with their decisions\footnote{Assume that there is a network where the agents are able to communicate or exchange their information. Here, the communication protocol is out of {}{the scope of this paper.}}. 
Relying on the updated actions, the agents will learn to achieve an acceptable solution together. 

In this paper, planning paths for multiple cameras at time step $t$ is modelled as an iterative game $\mathcal{G}_t$ where the players will play actions to gain their own long-run average benefit. The iterative game ${\mathcal{G}}_t$ consists of a finite set of players ${\mathcal{C}}$ as which cameras are regarded, a set of actions ${\mathcal{A}}_t$ and a set of utility functions of players ${\mathcal{U}}_t$ $({\mathcal{G}}_t=({\mathcal{C}},{\mathcal{A}}_t,{\mathcal{U}}_t))$. We denote by ${\mathcal{A}}_t={\mathcal{A}}_{t,1} \times {\mathcal{A}}_{t,2} \times \cdots \times  {\mathcal{A}}_{t,T}$ the set of all players' actions over $T$ iterations. Additionally, ${\mathcal{A}}_{t,\tau}={\mathcal{A}}^1_{t,\tau}\times {\mathcal{A}}^2_{t,\tau}\times \cdots \times {\mathcal{A}}^{|\mathcal{C}|}_{t,\tau}$ with $\tau=1,2,\dots,T$. Each player $c \in {\mathcal{C}}$ at iteration $\tau$ of time step $t$ has a finite set of actions ${\mathcal{A}}^c_{t,\tau}$ which consists of several levels of speed and moving angle. The number of speed levels and moving angles of $|\mathcal{C}|$ cameras are the same while they do not vary over time. We denote by ${\mathcal{U}}_t=\mathcal{U}_{t,1}\times \mathcal{U}_{t,2}\times \cdots \times \mathcal{U}_{t,T}$ the set of utility functions at time step $t$. Here, let $\mathcal{U}_{t,\tau} =\{u^1_{t,\tau},u^2_{t,\tau},\dots,u^{|\mathcal{C}|}_{t,\tau}\}$ define the set of utility functions of $|\mathcal{C}|$ cameras at iteration $\tau$ of time step $t$.

We denote by ${a}^c_{t,\tau}$ the action which camera $c$ performs at iteration $\tau$ of time step $t$ $({a}^c_{t,\tau} \in \mathcal{A}^c_{t,\tau})$. Here, a speed level $s^c_{t,\tau}$ and a moving angle $\phi^c_{t,\tau}$ of camera $c$ are included in an action ${a}^c_{t,\tau}$ $({a}^c_{t,\tau}=(s^c_{t,\tau},\phi^c_{t,\tau}))$. According to the objective function in problem~\eqref{origin problem}, when each camera $c$ plays action ${a}^c_{t,\tau}$, the utility function of camera $c$ is given by:
\begin{equation}
        \setlength\abovedisplayskip{0pt}
        \setlength\belowdisplayskip{0pt}
    \begin{aligned}
    u^c_{t,\tau}({a}^c_{t,\tau},{a}^{\overline{c}}_{t,\tau}) =  
        \begin{split}
            &-\Big[\Big(\text{Tr} \left(\doubletilde{{P}}^v_{t+1}\right) - \epsilon \Big)\\     &+\alpha_1 (D_{\min}-\hat{d}^{cv}_{t+1})(\hat{d}^{cv}_{t+1}-D_{\max})\\
            &+\alpha_2 \sum_{v'\in \mathcal{V}}(D_{\text{safe}}-\hat{d}^{cv'}_{t+1})
            \\
            &+\alpha_3\sum_{c\in \mathcal{C}}\sum_{c'\in\mathcal{C},c'\neq c} (D_{\text{safe}} - d^{cc'}_{t+1} )\Big],
        \end{split}                 
    \end{aligned}
    \label{eq:utility}
\end{equation}
where $v$ is the target assigned to camera $c$ using the sensor allocation process in section~\ref{sensor allocation}. Moreover, we denote by ${a}^{\overline{c}}_{t,\tau}$ the actions which the other cameras $\mathcal{C}\setminus \{c\}$ play at iteration $\tau$ of time step $t$. At each iteration $\tau$, given the other agents' actions ${a}^{\overline{c}}_{t,\tau}$, the agent $c$ 
picks an action $a^c_{t,\tau}$ until it achieves a maximum utility. Maximizing the sum of $|\mathcal{C}|$ agents' utilities, we will achieve the minimum value of the objective function in problem~\eqref{origin problem}.

According to~\cite{Aumann1987,Hart2000}, a game-based solution will converge at a set of equilibria where each player cannot 
improve its utility by changing its decision. Moreover, the equilibrium of our iterative game $\mathcal{G}_t$ is a correlated equilibrium (CE). We define by $\pi^c_{t,\tau}$ a probability distribution from which an agent $c$ selects an action $a^c_{t,\tau}$ in an action set $\mathcal{A}^c_{t,\tau}$. The probability distribution $\pi^c_{t,\tau}$ will be a CE if it holds true that 
\begin{equation}
        \setlength\abovedisplayskip{0pt}
        \setlength\belowdisplayskip{0pt}
    \sum_{a^{\overline{c}}_{t,\tau} \in \mathcal{A}^{\overline{c}}_{t,\tau}} \pi^c_{t,\tau} \big(a^c_{t,\tau},a^{\overline{c}}_{t,\tau}\big)
    \bigg[u^c_{t,\tau}({a'}^c_{t,\tau},a^{\overline{c}}_{t,\tau}) - u^c_{t,\tau}(a^c_{t,\tau},a^{\overline{c}}_{t,\tau}) \bigg] \leq 0,
\end{equation}
for all $a^{\overline{c}}_{t,\tau} \in \mathcal{A}^{\overline{c}}_{t,\tau}$ and for every action pair $a^c_{t,\tau},{a'}^c_{t,\tau} \in \mathcal{A}^c_{t,\tau}$. In a CE, each agent does not gain any benefits by choosing the other actions according to the probability distribution $\pi^c_{t,\tau}$ if the other agents do not change their decisions.

\begin{table}[t]
\caption{Experiment Settings.}
\begin{center}
\renewcommand{\arraystretch}{1.3}
\begin{tabular}{@{}lr@{}}\toprule
 \textbf{Parameter} & \textbf{Value} \\ 
 \hline
 \multicolumn{2}{c}{\textbf{Stationary radar}}\\
 $p^r$ & $13$ \\
 \hline
 \multicolumn{2}{c}{\textbf{Automatic identification systems (AIS)}}\\
 $\sigma^{\text{AIS}}$ & $10^3$ \\
 \hline
 \multicolumn{2}{c}{\textbf{Mobile cameras}}\\
 $p^c$ & $13$ \\
 Number of mobile cameras $|\mathcal{C}|$& $[2;30]$ \\
 HFOV of each camera $c$ & $3 \degree$ \\
 Detection range of each camera $c$ & $16$ km \\
 $\Delta t$ & $1$ s \\ 
 $H$ & $[1;5]$ time steps \\
 $\alpha_1$, $\alpha_2$, $\alpha_3$ & 1 \\
 $D_{\min}$ & $80$ m \\
 $D_{\max}$ & $16$ km \\
 $D_{\text{safe}}$ & $100$ m \\
 $s_{\min}$ & $0$ knot ($= 0$ m/s) \\
 $s_{\max}$ & $18$ knots ($\approx 9$ m/s)\\
 $\Phi_{\max}$ & $180 \degree$  \\
 $T$ & $[10;20]$ \\
 $\epsilon$ & $10^4$ \\
 $\lambda$ & $0.5$ \\
 Number of cameras' speed levels $|\mathcal{L}_s|$ & $[2;10]$ \\
 Number of cameras' heading angles $|\mathcal{L}_m|$ & $[4;20]$ \\
 \hline
 \multicolumn{2}{c}{\textbf{Mobile targets}}\\
 Number of targets $|\mathcal{V}|$ & $[3;100]$ \\
 Speed of each target $v$ & $[0;18]$ knots ($\approx [0;9]$ \text{m/s})\\
 $\sigma_v$ & $3$ \\
 \bottomrule
\end{tabular}
\end{center}
\label{table:settings}
\end{table}

In the proposed distributed multi-agent path planning algorithm, we employ a learning process, regret-matching procedure. According to this procedure, each agent is allowed to individually compute the probability of its action which is proportional to the regrets for not selecting the other actions. Specifically, at an iteration $\tau$ of time step $t$, for any two actions $j,k \in \mathcal{A}^c_{t,\tau}$ and $j \neq k$, the cumulative regret of an agent $c$ for not selecting action $k$ instead of action $j$ up to iteration $\tau$ is given by:
\begin{equation}
        \setlength\abovedisplayskip{0pt}
        \setlength\belowdisplayskip{0pt}
    \Hat{D}^c_{t,\tau}(j,k)=\frac{1}{\tau}\sum_{\Hat{\tau}=1}^\tau \mathbbm{1}(a^c_{t,\Hat{\tau}}=j)\{
        u^c_{t,\Hat{\tau}}(k,a^{\overline{c}}_{t,\Hat{\tau}})-u^c_{t,\Hat{\tau}}(j,a^{\overline{c}}_{t,\hat{\tau}})\},
\end{equation}
where $\mathbbm{1}(.)$ is an indicator function. If the value of the regret is negative, the agent $c$ will not regret when playing action $j$ instead of action $k$. Here, we can express the cumulative regret recursively as follows:
\begin{equation}
        \setlength\abovedisplayskip{0pt}
        \setlength\belowdisplayskip{0pt}
    \Hat{D}^c_{t,\tau}(j,k)=\left(1-\frac{1}{\tau}\right)\Hat{D}^c_{t,\tau-1}(j,k)+\frac{1}{\tau}D^c_{t,\tau}(j,k),
    \label{eq:origin_probability}
\end{equation}
where ${D}^c_{t,\tau}(j,k)= \mathbbm{1}(a^c_{t,\tau}=j)\{
        u^c_{t,\tau}(k,a^{\overline{c}}_{t,\tau})-u^c_{t,\tau}(j,a^{\overline{c}}_{t,\tau})\}$ defines the instantaneous regret when agent $c$ has not played action $k$ instead of action $j$. Through Eq.~\eqref{eq:origin_probability}, it is noteworthy that there is a significant influence of the outdated regret $\Hat{D}^c_{t,\tau-1}(j,k)$ on the instantaneous regret $\Hat{D}^c_{t,\tau}(j,k)$. It causes a slow convergence speed for our proposed path planning algorithm. Meanwhile, due to a high mobility environment like maritime environment, our proposed path planning algorithm is required to converge at a CE with a high speed. Therefore, like~\cite{Long2022}, we employ a forgetting factor for updating $ \Hat{D}^c_{t,\tau}(j,k)$ in the following.
\begin{equation}
        \setlength\abovedisplayskip{0pt}
        \setlength\belowdisplayskip{0pt}
    \Hat{D}^c_{t,\tau}(j,k)=\lambda \cdot \Hat{D}^c_{t,\tau-1}(j,k)+(1-\lambda)D^c_{t,\tau}(j,k),
    \label{eq:probability}
\end{equation}
where $\lambda \in [0;1]$ is the forgetting factor. Here, the influence of outdated regret values over the instantaneous regret will be regulated by $\lambda$. In addition, {}{the advantages of forgetting factor $\lambda$} have been demonstrated in our previous work~\cite{Long2022}. As a result, the probability that agent $c$ plays an action $k$ in the next iteration at time step $t$ is computed by:
\begin{equation}
        \setlength\abovedisplayskip{0pt}
        \setlength\belowdisplayskip{0pt}
    \pi^c_{t,\tau+1}(k) = \frac{1}{\mu}  \max\{\Hat{D}^c_{t,\tau}(j,k),0\} \quad \forall k \neq j, 
    \label{eq:updated_probability}
\end{equation}
where $\mu$ is a normalizing factor to ensure that the probability $\pi^c_{t,\tau+1}$ sums to $1$. Then, the probability that action $j$ is chosen by agent $c$ in the next iteration is given by:
\begin{equation}
        \setlength\abovedisplayskip{0pt}
        \setlength\belowdisplayskip{0pt}
    \begin{aligned}
        & \pi^c_{t,\tau+1}(j)=1-\sum_{k\neq j, k \in \mathcal{A}^c_{t,\tau}} \pi^c_{t,\tau+1}(k).
    \end{aligned}
    \label{eq:rml}
\end{equation}

 \begin{figure}[!tbp]
        \centering
        \includegraphics[width=0.9
\columnwidth]{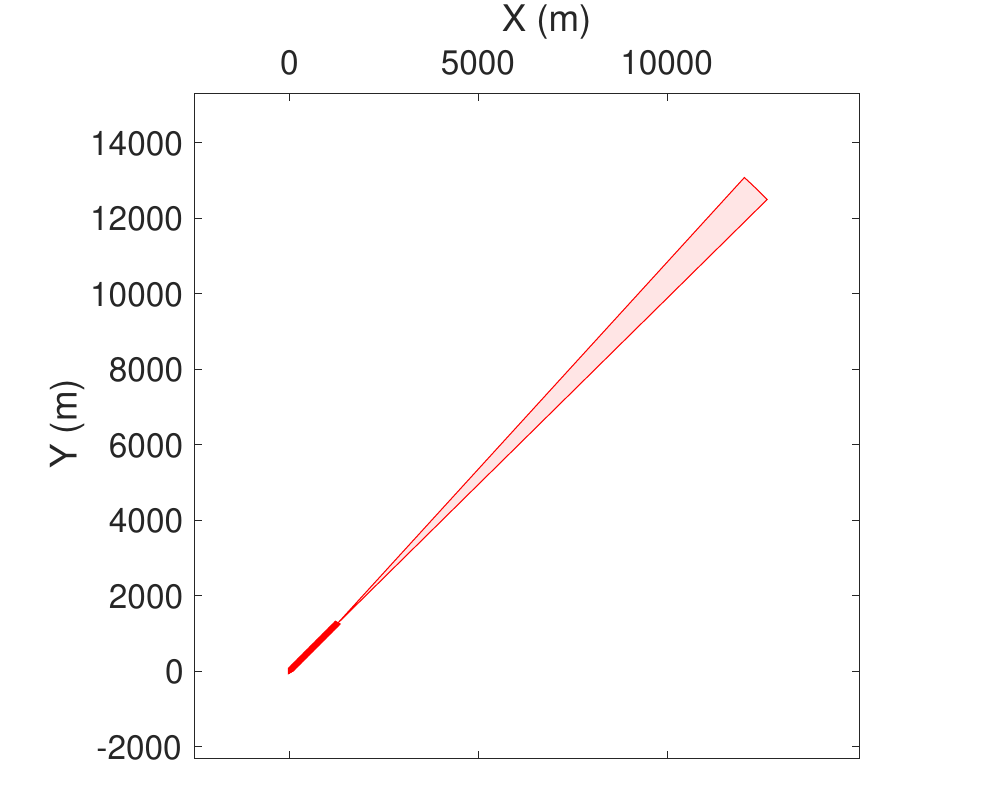}
        \captionsetup{justification=centering}
        \caption{{}{The camera trajectory} is shown after moving from $(0,0)$ to $(1260,1260)$. Also shown in the FOV at location $(1260,1260)$. Here, the camera FOV is $3 \degree$ while the camera detection range is $16$~km, similar to the VIGY observer made by SAFRAN~\cite{Safran2023}.}
        \label{fig:camera_FOV}
\end{figure} 

\section{Computational Study} \label{sec:experiment results}
\subsection{Experiment settings}
\begin{figure}[!tbp]
    \begin{minipage}[t]{0.5\textwidth}
        \centering
        \includegraphics[width=1\columnwidth]{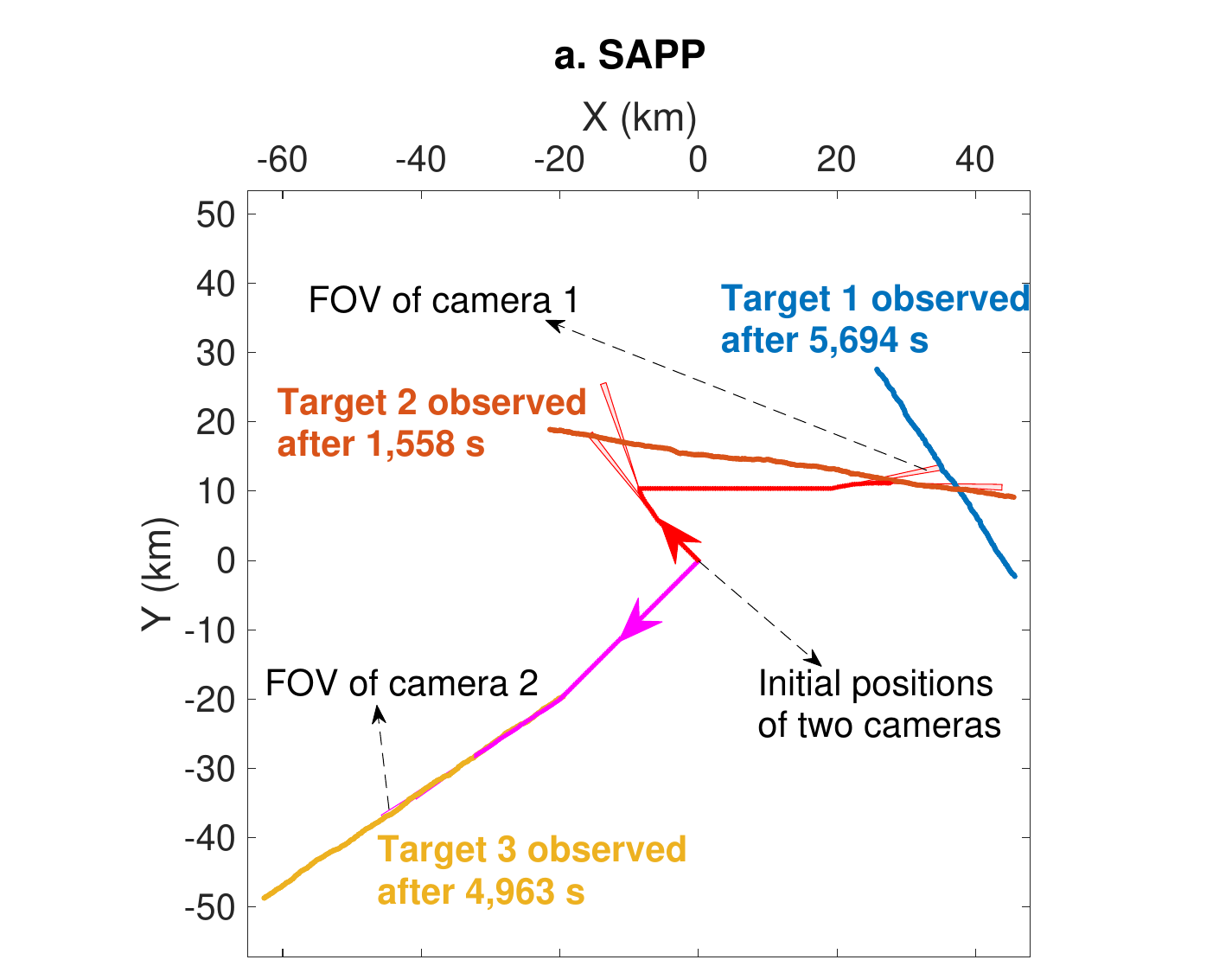}
        \vspace{3mm}
    \end{minipage}\vspace{-6mm}
    \begin{minipage}[t]{0.5\textwidth}
    \centering
        \includegraphics[width=1
\columnwidth]{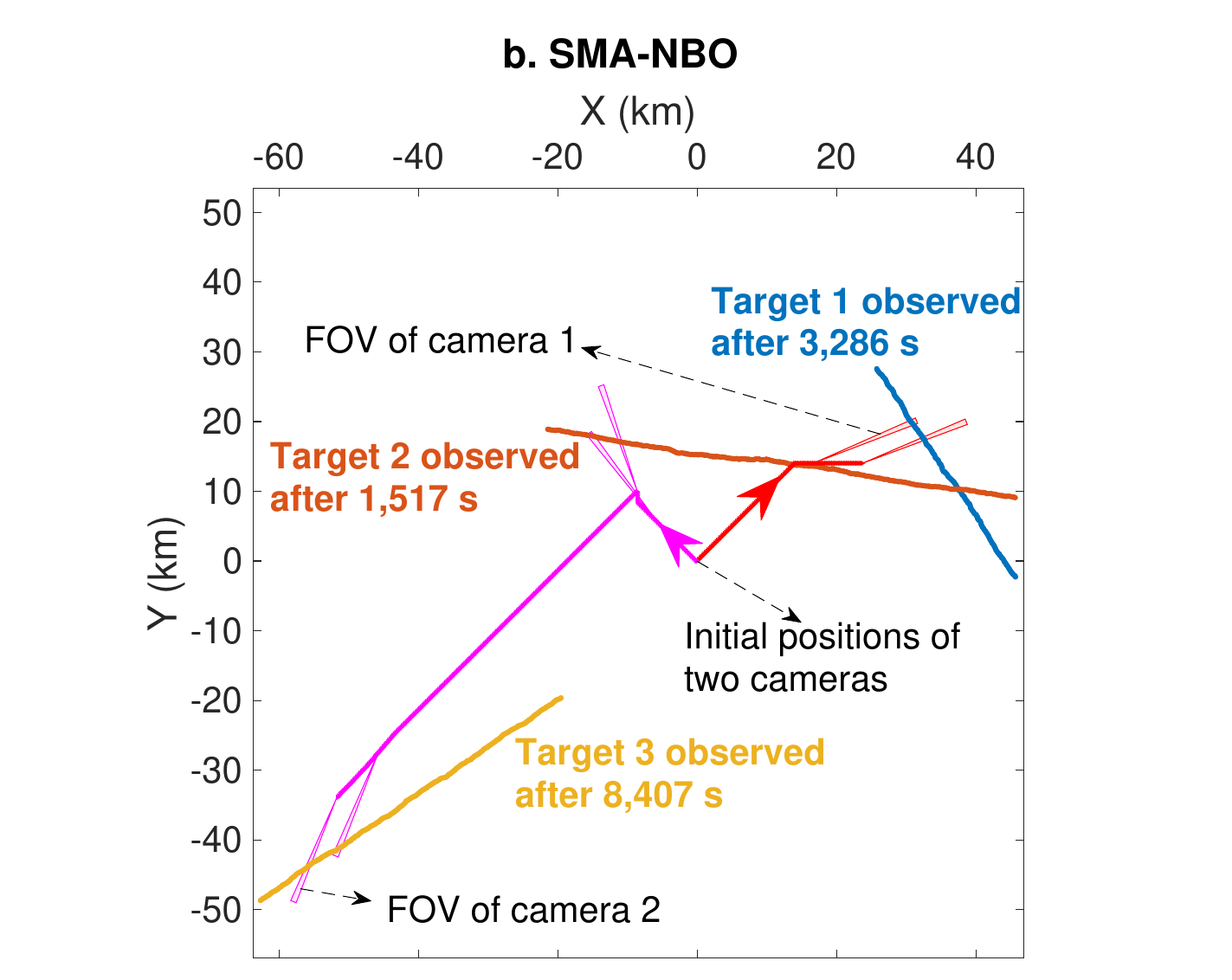}
    \end{minipage}  
    \captionsetup{justification=centering}
    \caption{{}{Paths (in magenta and red) planned over $1$ time step for $2$ cameras to observe $3$ targets.}}
    \label{fig:2cams_3targets_paths}
\end{figure}

To evaluate the performance and efficiency of SAPP, we conduct computer-based experiments with practical settings. In the experiments, one static radar and AIS always update us with their observations of target positions at the end of each time step. The radar is installed at the position of $(95~\text{km},-95~\text{km})$. Unlike the radar, cameras are able to move around to observe targets, and they are required to observe these targets once. The camera positions are initially set as $(0,0)$ while the positions of targets are randomly distributed between $15~\text{km}$ and $30~\text{km}$ away from the origin. In addition, the targets do not reside within any cameras' FOV. The camera FOV is illustrated in Fig.~\ref{fig:camera_FOV}. Similar to~\cite{Nguyen2019}, the uncertainty in measurements of radar $r$ and camera $c$ is given by $\sigma^{rv}_t=d^{rv}_t \cdot \frac{p^r}{100}$ and $\sigma^{cv}_t=d^{cv}_t \cdot \frac{p^c}{100}$, respectively. Here, $d^{rv}_t$ and $d^{cv}_t$ are the distance between the sensors and target $v$ at time step $t$, respectively. The values of all parameters in our experiments are based on \cite{Nguyen2019,Safran2023,Ragi2013,Imo2023}, and they are listed in Table~\ref{table:settings}. 

\begin{figure}[!tbp]
    \begin{minipage}[t]{0.5\textwidth}
        \centering
        \includegraphics[width=0.9\columnwidth]{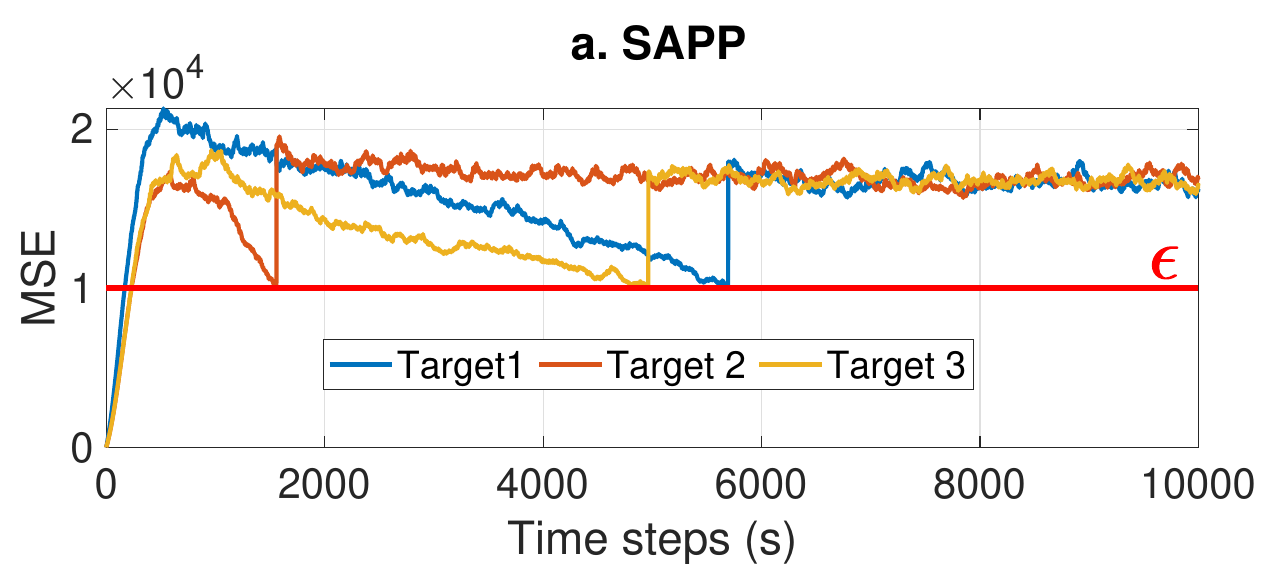}
        \vspace{3mm}
    \end{minipage}
    
    \begin{minipage}[t]{0.5\textwidth}
    \centering
        \includegraphics[width=0.9
\columnwidth]{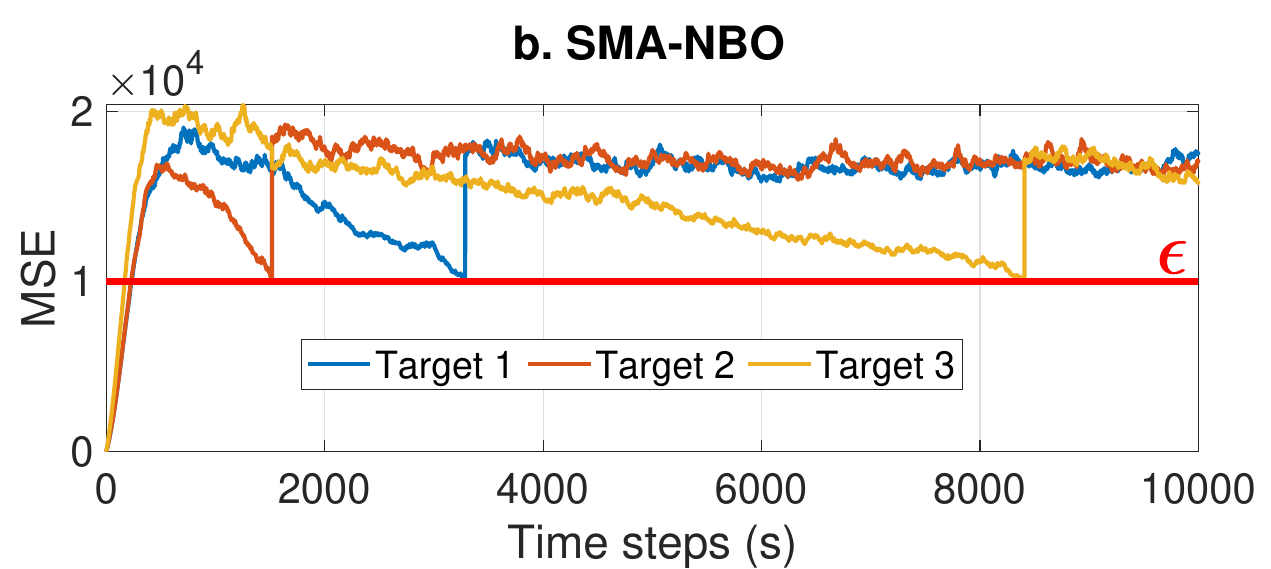}
    \end{minipage}  
        \captionsetup{justification=centering}
        \caption{{}{The MSE for each target in the scenario where $2$ cameras are trying to observe $3$ targets.}}
    \label{fig:2cams_3targets_MSE}
\end{figure} 

To demonstrate how SAPP works, we compare our approach with the most relevant work in literature. Specifically, the related work~\cite{Li2022}{}{, Sequential
Multi-Agent Planning with Nominal Belief-State Optimization (SMA-NBO),} has been recently published in 2022, and it aims to minimize the total MSE. In SMA-NBO, the targets leave the limited FOV of sensors mounted on UAVs at the end of each time step. Thus, when the targets are outside the sensors' FOV, a target whose MSE is the greatest will be assigned to the nearest UAV. Then, using a greedy algorithm, {SMA-NBO} generates the trajectories for multiple UAVs sequentially in a centralized manner to guarantee that each UAV covers the nominal mean position of the assigned target. Here, we re-implement {SMA-NBO} in our maritime scenario without any changes.

We evaluate the performance of the two approaches in terms of achieving situational awareness as quickly as possible, and using the 
following two metrics.  
\begin{enumerate}
    \item \textbf{Mean squared error (MSE)}: which represents situational awareness improvement, and is determined through $\text{Tr} \left(\protect\doubletilde{{P}}^v_{t+1}\right)$ in the objective function~\eqref{objective function problem},
    \item \textbf{Observation duration}: which is computed from when we start conducting an experiment to when the final target in the experiment is observed.
\end{enumerate}
\begin{figure}[!tbp]
    \begin{minipage}[t]{0.5\textwidth}
        \centering
        \includegraphics[width=1\columnwidth]{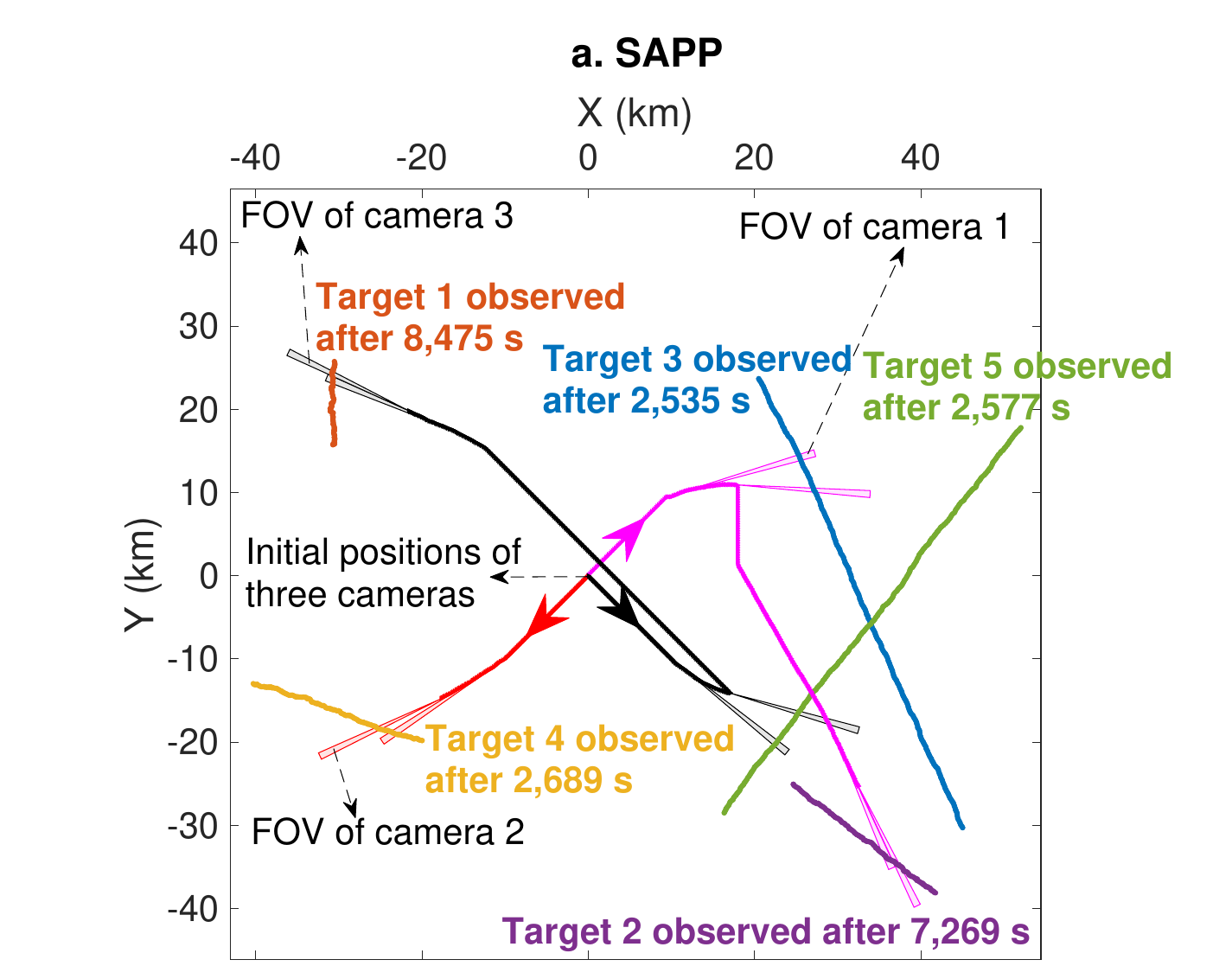}
        \vspace{3mm}
    \end{minipage}\vspace{-6mm}
    \begin{minipage}[t]{0.5\textwidth}
    \centering
        \includegraphics[width=1
\columnwidth]{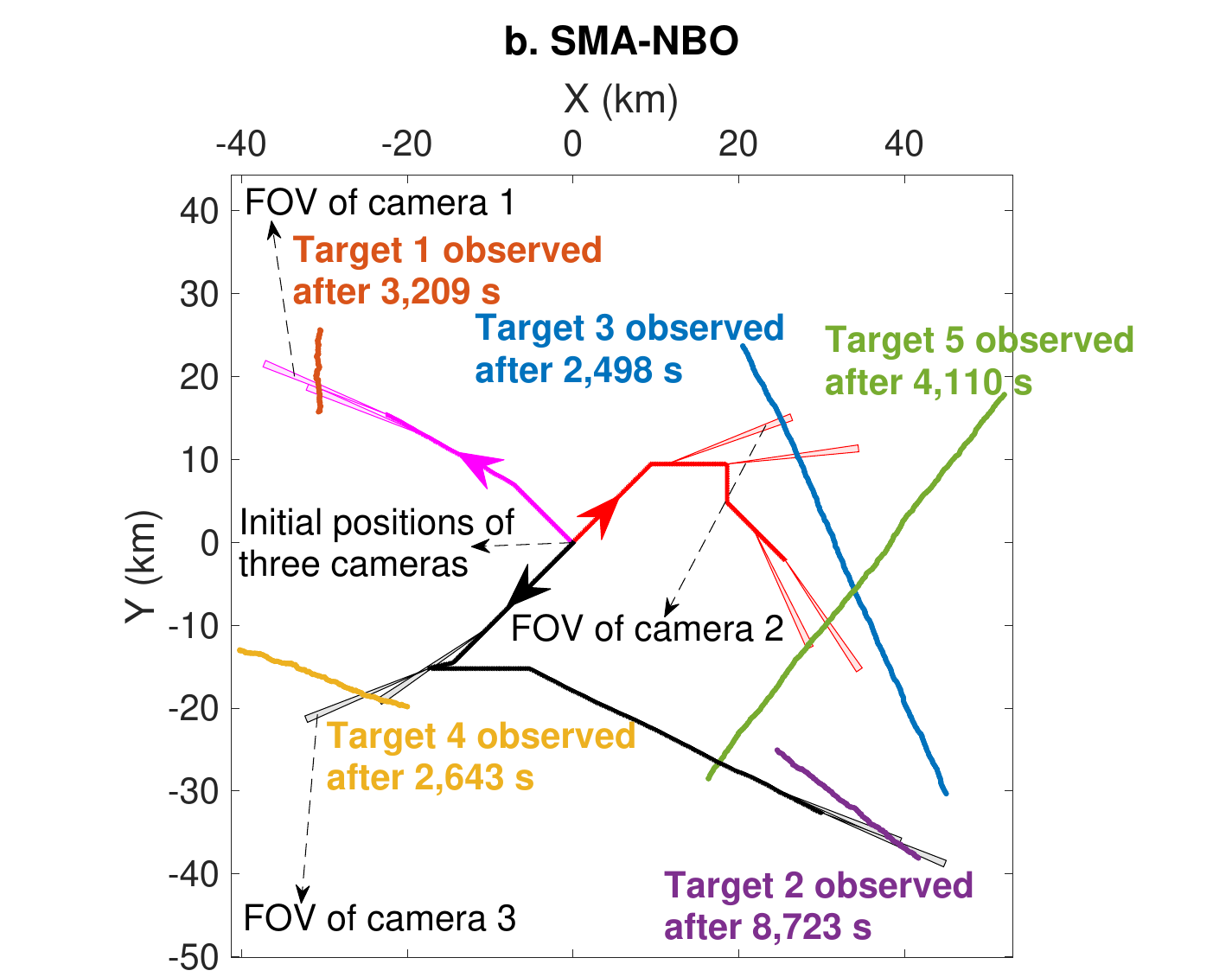}
    \end{minipage}    
        \captionsetup{justification=centering}
        \caption{{}{Paths (in red, magenta and black) planned over $5$ time steps for $3$ cameras to observe $5$ targets
        .}}
        \label{fig:3cams_5targets_paths}
\end{figure} 
\begin{figure}[h]
    \begin{minipage}[t]{0.5\textwidth}
        \centering
        \includegraphics[width=0.9\columnwidth]{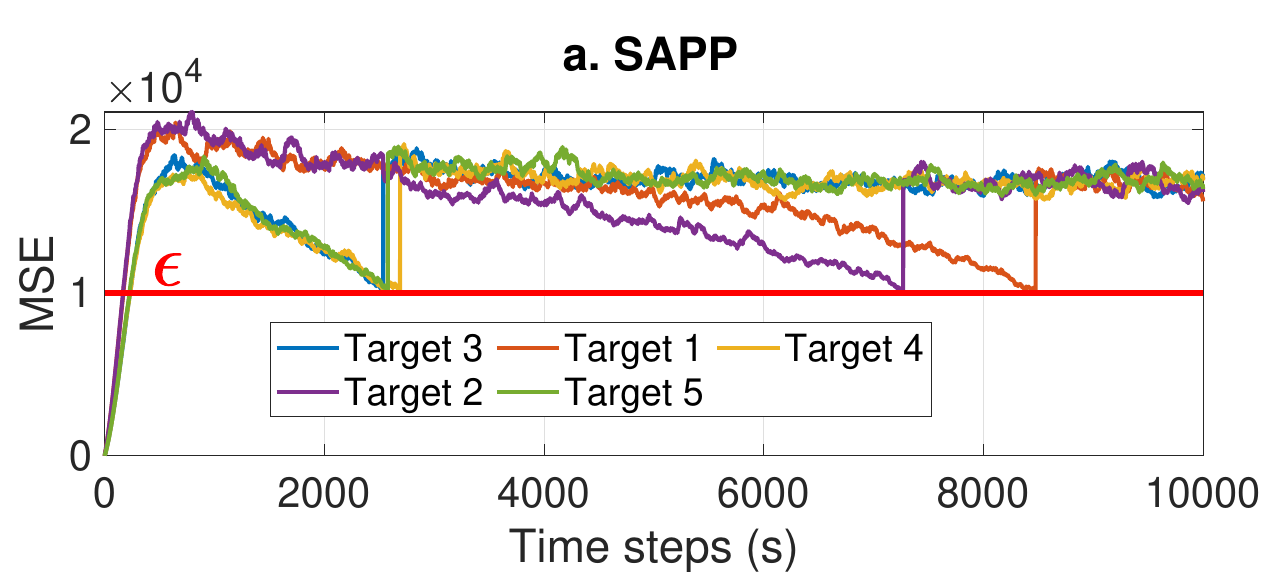}
        \vspace{3mm}
    \end{minipage}
    
    \begin{minipage}[t]{0.5\textwidth}
    \centering
        \includegraphics[width=0.9
\columnwidth]{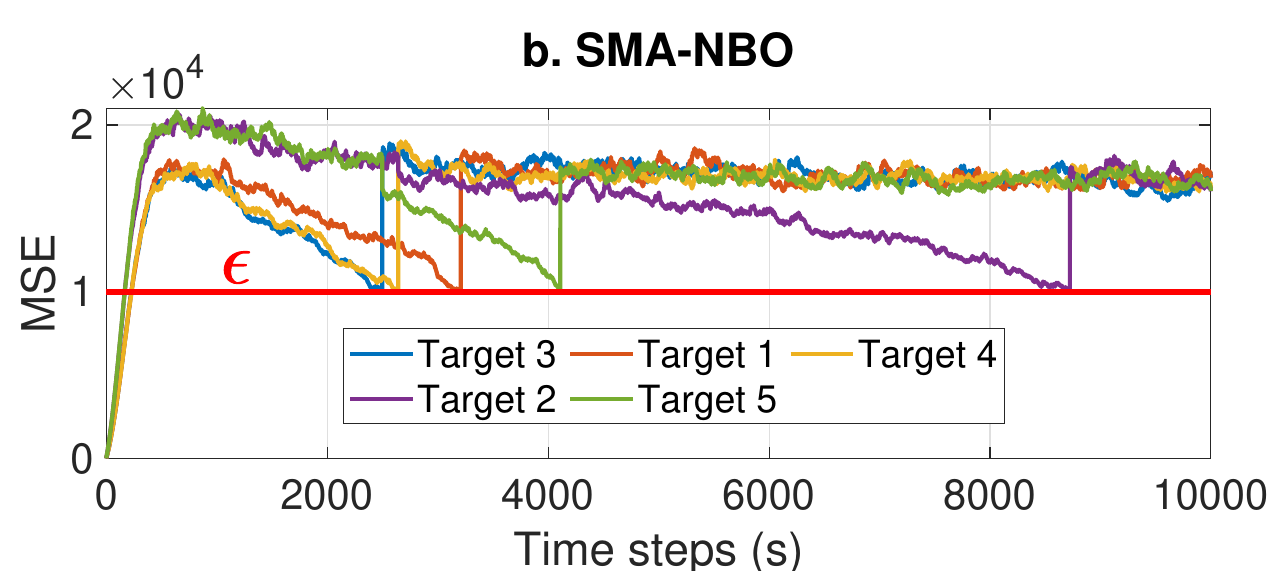}
    \end{minipage}  
        \captionsetup{justification=centering}
        \caption{{}{The MSE for each target in the scenario where $3$ cameras are trying to observe $5$ targets.}}
    \label{fig:3cams_5targets_MSE}
\end{figure}

{Our scenarios extend comparable ones from the literature.} SMA-NBO of \cite{Li2022} considered six small scenarios, i.e. from 
$<|\mathcal{C}|=2,|\mathcal{V}|=3,H=1>$ to $<|\mathcal{C}|=3,|\mathcal{V}|=5,H=5>$, and we extend these scenarios by several orders of magnitude by considering up to $|\mathcal{C}|=30$ cameras and $|\mathcal{V}|=100$ targets. 
{}{Also, we note that, for each pair of scenarios that differ only in the planning horizon $H$, (e.g. Scenarios $11$ and $12$), the trajectories of the targets are the same (e.g. the trajectories in instance $55$ of Scenario $11$ are identical to those in instance $55$ of Scenario $12$).}

Furthermore, to fairly compare the two works, we always set up the same scenario with the same parameter settings (including the same random number generator seed) when running the experiments. Our implementation of our algorithm and {SMA-NBO} are using only one CPU core. The experiments are run on MonARCH (Monash Advanced Research Computing Hybrid)\footnote{\url{https://docs.monarch.erc.monash.edu.au/index.html}}.




\subsection{Results}

\begin{figure}[!tbp]
    \begin{minipage}[t]{0.46\textwidth}
        \centering
        \includegraphics[width=1\columnwidth]{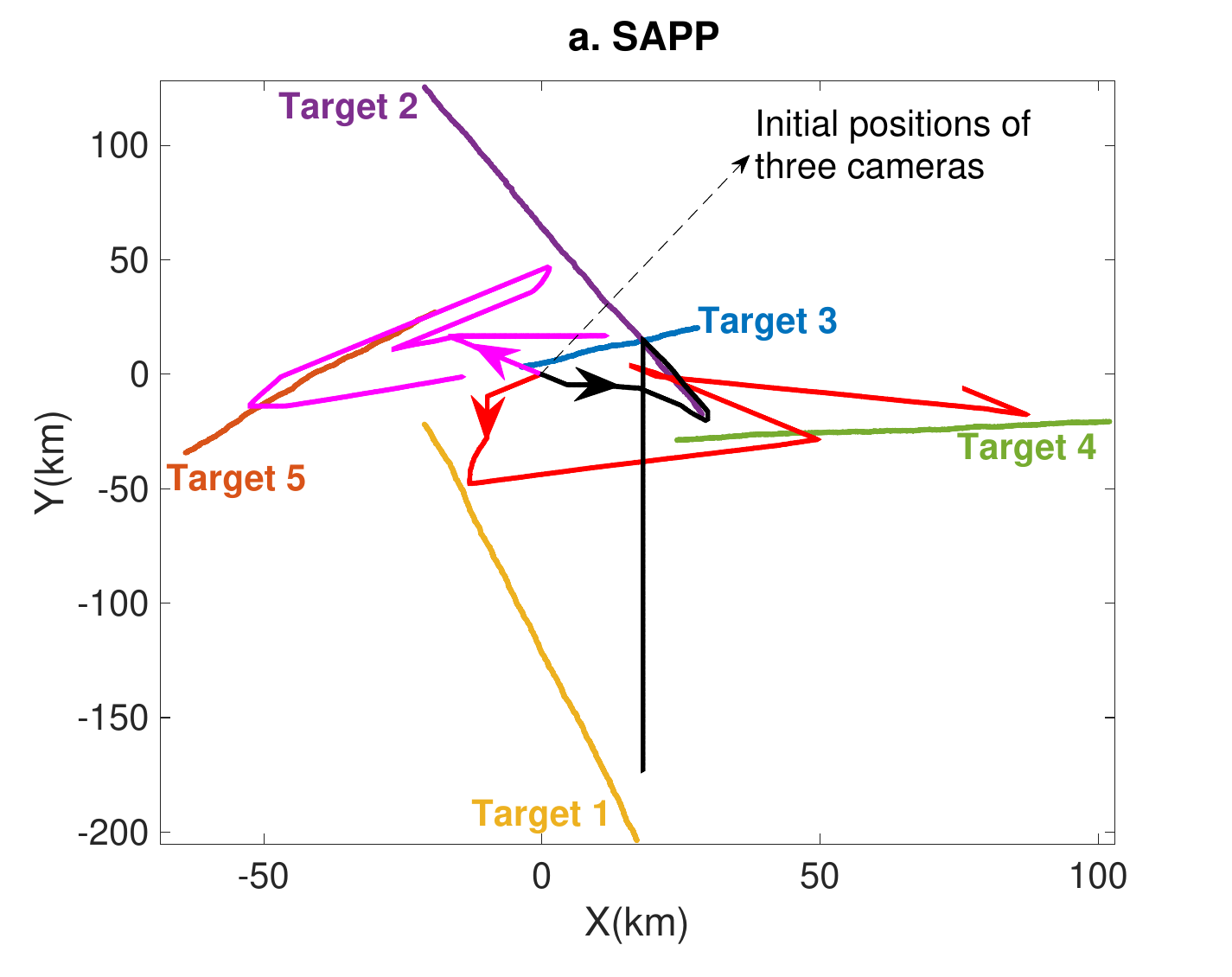}
        \vspace{3mm}
    \end{minipage}\vspace{-7mm}  
    \begin{minipage}[t]{0.46\textwidth}
    \centering
        \includegraphics[width=1
\columnwidth]{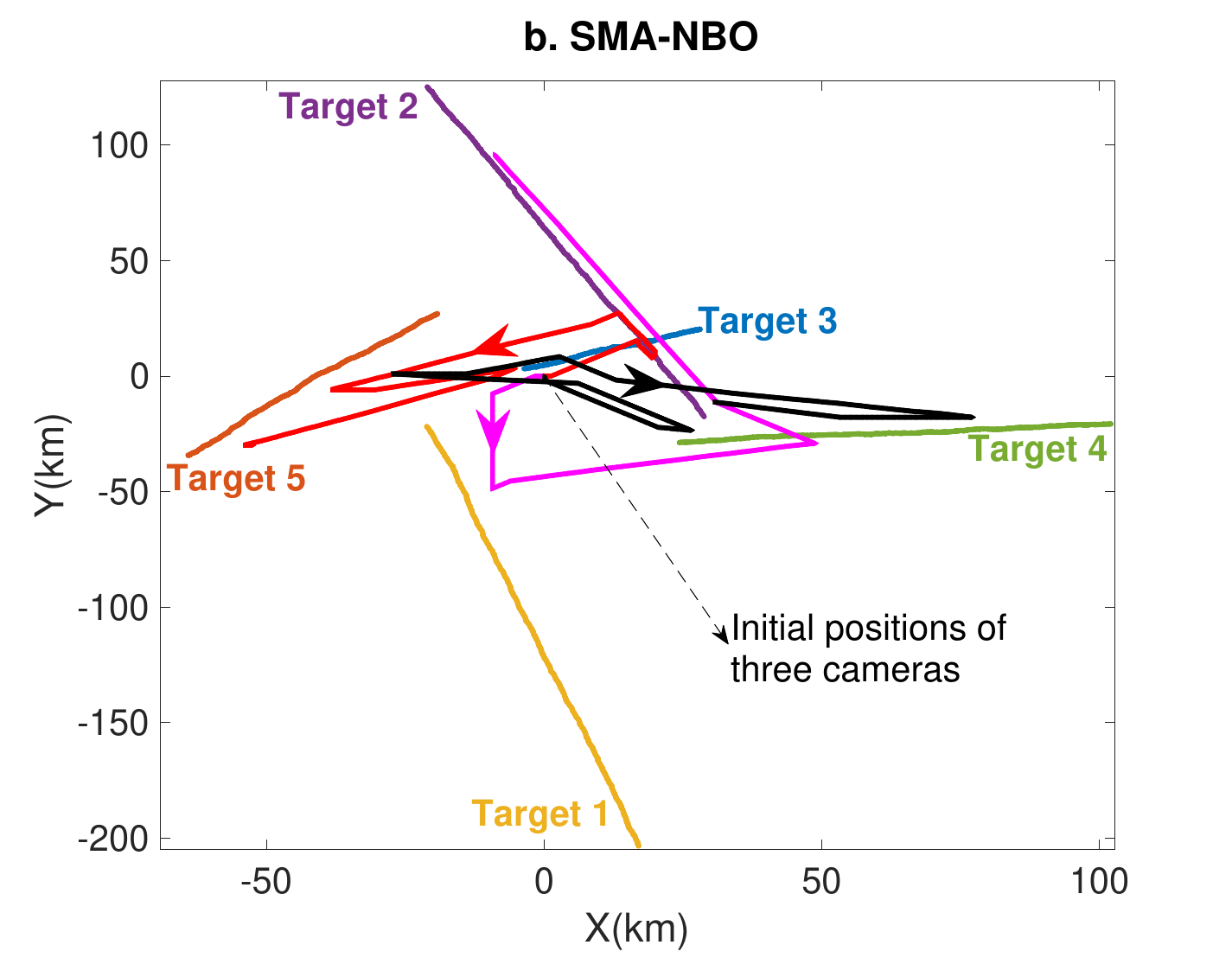}
    \end{minipage}    
        \captionsetup{justification=centering}
        \caption{{}{Paths (in red, black and magenta) planned for $3$ cameras to observe $5$ targets many times using SAPP and SMA-NBO.}}
        \label{fig:3cams_5targets_times}
\end{figure} 
\begin{figure}[h]
    \begin{minipage}[t]{0.48\textwidth}
        \centering
        \includegraphics[width=1\columnwidth]{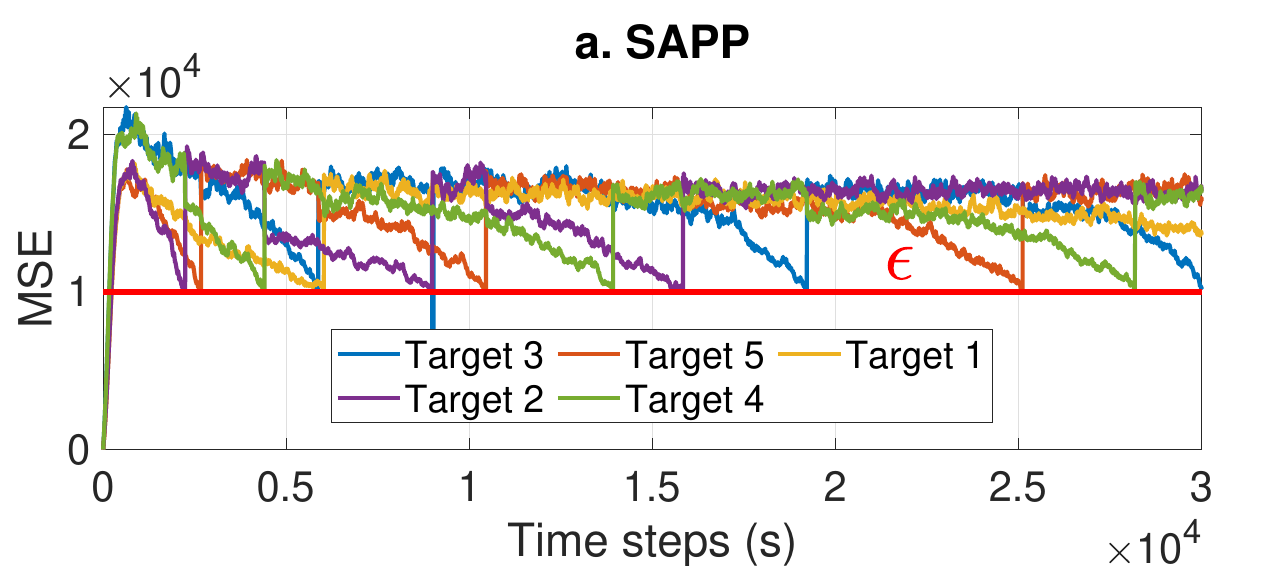}
        \vspace{3mm}
    \end{minipage}\vspace{-4mm}
    \begin{minipage}[t]{0.48\textwidth}
    \centering
        \includegraphics[width=1
\columnwidth]{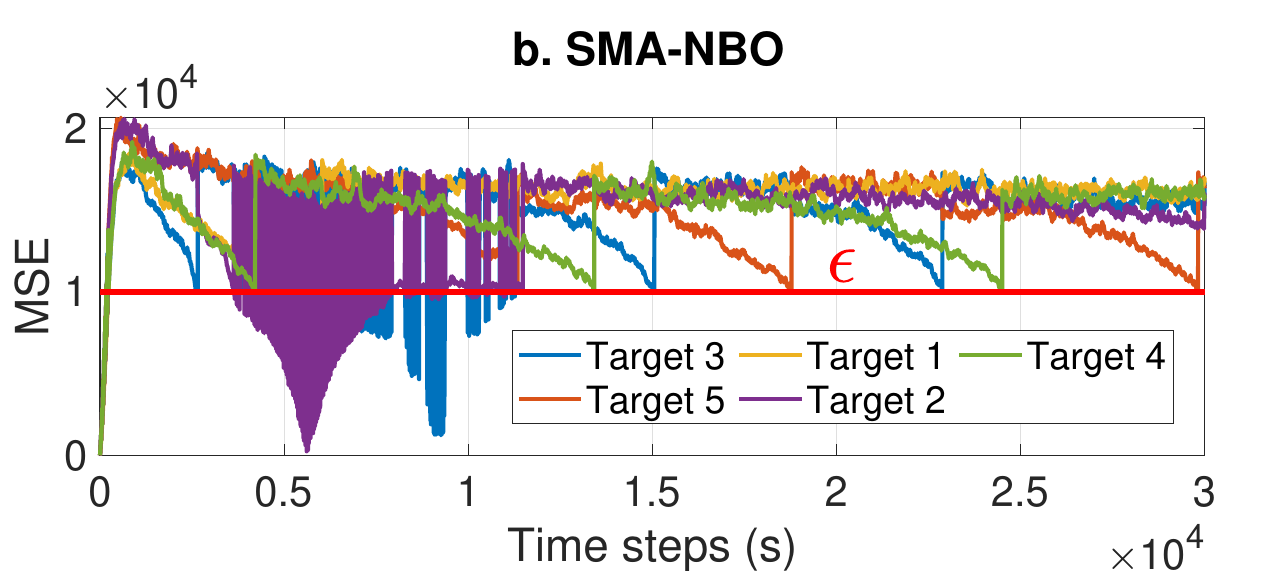}
    \end{minipage}  
        \captionsetup{justification=centering}
        \caption{{}{MSE of $5$ targets minimized by SAPP and SMA-NBO using $3$ cameras.}}
    \label{fig:3cams_5targets_MSE_times}
\end{figure} 

In Figs.~\ref{fig:2cams_3targets_paths} and~\ref{fig:3cams_5targets_paths}, we show for two scenarios the trajectories produced by SAPP and contrast the results with those achieved by {SMA-NBO}: 2 cameras aim to observe 3 targets in the first scenario, and 3 cameras aim to observe 5 targets in the second scenario. We can see that our approach achieves situational awareness (as measured by the time needed to observe all targets at least once) much quicker: $5,694$~s and $8,407$~s versus $8,475$~s and $8,723$~s. 
That is because our camera allocation depends on not only the distance between the cameras and the targets but also how long the targets stay unobserved. In addition, our distributed path planning algorithm guarantees a solution, which is a correlated equilibirum solution 
for all the cameras. {}{In contrast, SMA-NBO relies on the target MSE and the distance between cameras and targets to assign the cameras to the targets. Moreover, the greedy algorithm in SMA-NBO is not able to maintain a convergence at a good solution for planning camera trajectories.} Figs.~\ref{fig:2cams_3targets_MSE} and~\ref{fig:3cams_5targets_MSE} complement the observation by focusing on the MSE: they show that both the approaches are able to enhance situational awareness using the cameras. Initially, due to a high uncertainty in observations from radar $r$ and AIS, the MSE of the two approaches remains high when we only rely on these observations. By contrast, the MSE is reduced to the pre-given threshold level $\epsilon$ when the targets reside within the cameras' FOV and are perceived by the cameras. That is because the cameras are able to provide us measurements with a lower uncertainty. 

\begin{table*}[t]
\centering
\caption{Comparison between SAPP and~{}{SMA-NBO} in terms of observation duration and execution time
when varying the number of cameras and targets, and the length of time horizon.}

\renewcommand{\arraystretch}{1.5}
\begin{tabular}
{@{}P{1cm}P{1cm}P{1cm}P{1cm}P{1.6cm}P{1.6cm}P{1.5cm}P{1.6cm}P{1.6cm}P{1.5cm}@{}}\toprule 
 \multirow{2}{*}{Scenarios} 
 & \multirow{2}{*}{$|\mathcal{C}|$} 
 & \multirow{2}{*}{$|\mathcal{V}|$} 
 & \multirow{2}{*}{$H$}  
 &\multicolumn{3}{c}{\textbf{SAPP}} 
 & \multicolumn{3}{c}{{}{SMA-NBO}} \\
  \cline{5-10}
    & & & & Average observation duration 
    & Average execution time (s) 
    & Failed runs$^7$
    & Average observation duration 
    & Average execution time (s) 
    & Failed runs \\
  \hline
  $1$ & \multirow{2}{*}{$2$} & \multirow{2}{*}{$3$} & $1$ & $\mathbf{7.06\times 10^3}$ & $0.020$  & $0$ & $7.13\times 10^3$ & $0.004$ & $1$ \\
  $2$ & & & $5$ & $7.14 \times 10^3$ & $0.093$ & $0$ & $\mathbf{7.12\times 10^3}$ & $0.001$ & $1$ \\
  \hline
  $3$ & \multirow{2}{*}{$3$} & \multirow{2}{*}{$4$} & $1$ & $6.93 \times 10^3$ & $0.027$ & $2$ & $\mathbf{6.65\times 10^3}$ & $0.006$ & $1$ \\
  $4$ & & & $5$ & ${6.92 \times 10^3}$ & $0.125$ & $2$ & $\mathbf{6.65\times 10^3}$ & $0.013$ & $1$\\
  \hline
  $5$ & \multirow{2}{*}{$3$} & \multirow{2}{*}{$5$} & $1$ & $\mathbf{7.97\times 10^3}$ & $0.036$ & $2$ & $8.91\times 10^3$ & $0.007$ & $3$ \\
  $6$ & & & $5$ & $\mathbf{8.03\times 10^3}$ 
   & $0.161$ & $2$ & $8.89\times 10^3$ 
   & $0.017$  & $3$ \\
  \hline
  $7$ &\multirow{2}{*}{$3$} & \multirow{2}{*}{$10$} & $1$ & $\mathbf{1.15 \times 10^4}$ & $0.065$ & $24$ & $1.37\times 10^4$ & $0.013$ & $49$ \\
  $8$ & & & $5$ & $\mathbf{1.15 \times 10^4}$ & $0.294$ & $23$ & $1.37 \times 10^4$ & $0.032$ & $49$\\
  \hline
    $9$ & \multirow{2}{*}{$30$} & \multirow{2}{*}{$50$} & $1$ 
    & $\mathbf{1.12 \times 10^4}$ 
    &   $1.170$
    & $16$
    &   $1.27 \times 10^4$
    &   $0.397$
    & $20$
    \\
   $10$ & &  & $5$ 
   & $\mathbf{1.15\times 10^4}$
   & $4.636$
   & $22$
   & $1.26\times 10^4$
   & $0.718$
   & $19$
   \\
   \hline
  $11$ & \multirow{2}{*}{$30$} & \multirow{2}{*}{$100$} & $1$ 
  & $\mathbf{1.40 \times 10^4}$ 
  & $1.987$
  & $54$
  & $1.50\times 10^4$ 
  & $0.759$ 
  & $93$ \\ 
  $12$ & & & $5$ 
  & $\mathbf{1.39\times 10^4}$ 
  & $7.973$ 
  & $57$
  & $1.50 \times 10^4$ 
  & $1.483$
  & $95$ \\
   \bottomrule
\end{tabular}
\label{table:comparison}\vspace{1mm}\\
\footnotesize{$^7$Failed runs refer to the number of runs (out of $100$) in which an approach failed to observe all targets in time (within $15,000$ time steps).}
\end{table*}

{}{Exemplarily, we show in Fig.~\ref{fig:3cams_5targets_times} the ``continuous observation'' of two approaches, i.e. we do not stop the experiment once all targets are observed at least once. As a result, the MSE of each target is reduced to the threshold level many times, as shown in Fig.~\ref{fig:3cams_5targets_MSE_times}. It demonstrates that SAPP and SMA-NBO are able to observe the targets continuously; however, they plan different trajectories for the cameras. Different from SAPP, one target in SMA-NBO might be looked at more times than the others although it has just been observed. That is because allocating cameras to targets in SMA-NBO depends on the target MSE instead of how long the targets have not been observed. Consequently, the MSE of target~$2$ is minimized many more times than that of the others, as shown in Fig.~\ref{fig:3cams_5targets_MSE_times}.
} 


\begin{figure}[h]
        \centering
        \includegraphics[width=1.1\columnwidth]{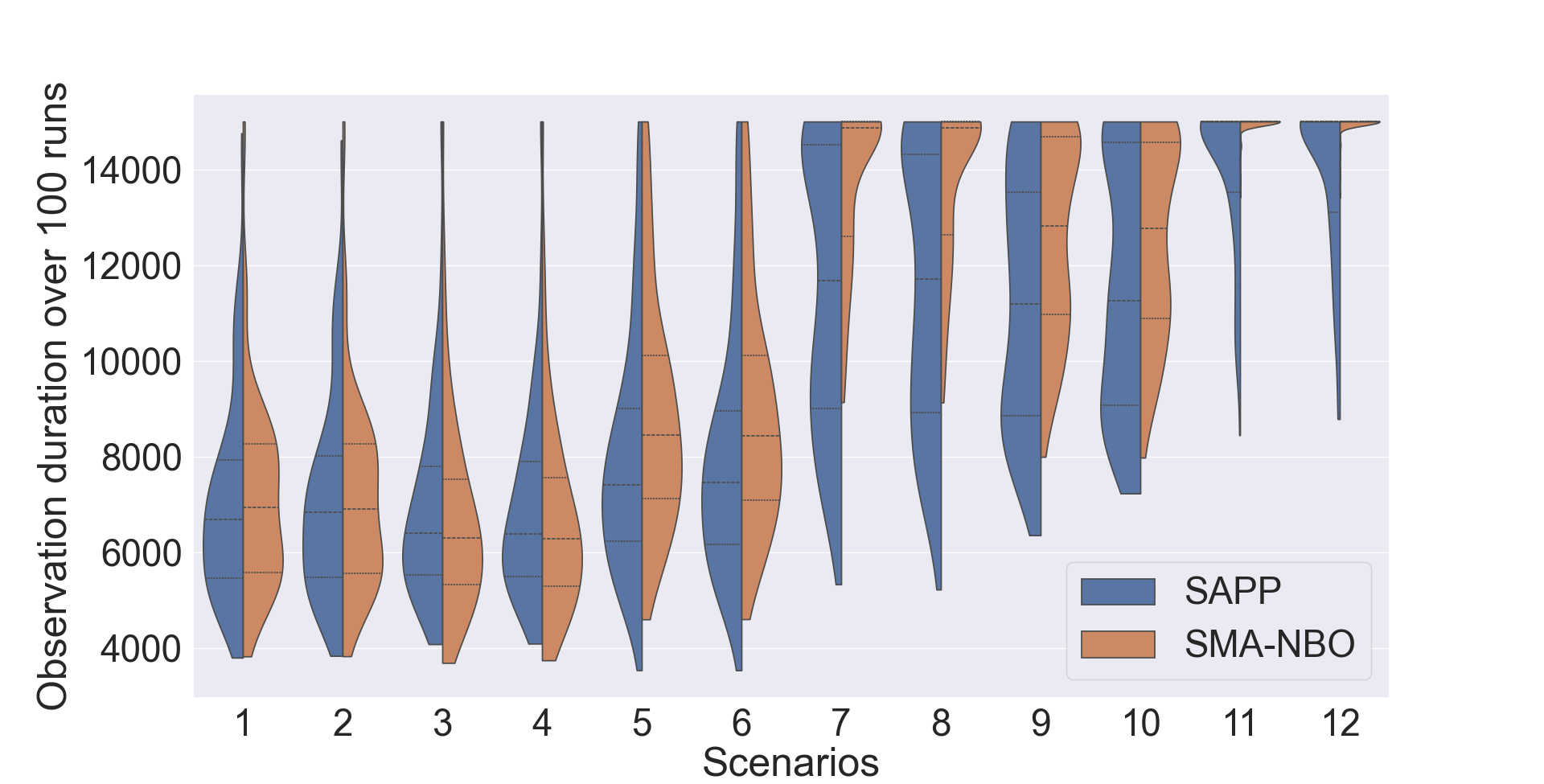}
        \captionsetup{justification=centering}
        \caption{{}{Violin plots for the comparison between SAPP and~{}{SMA-NBO} in terms of observation duration. 
        The violins indicate the actual distribution of the results over $100$ instances per~scenario. 
        }}
    \label{violin_plot}\vspace{3mm}
\end{figure}

{}{Table~\ref{table:comparison} and Figure~\ref{violin_plot} present how SAPP and~{}{SMA-NBO} perform in different scenarios where we vary the camera number $|\mathcal{C}|$, the target number $|\mathcal{V}|$ and the time horizon length $H$. 
Shown are (per row) the results for $100$ instances; each instance's budget is $15,000$ time steps. 
The ``average observation duration'' refers to our objective: the average number of time steps needed (across all $100$ instances) to observe all targets once; lowest averages are highlighted in bold. The ``average execution time (s)'' refers to the computational cost in seconds per time step.} 


{}{In almost all scenarios, SAPP observes all targets more quickly than~{}{SMA-NBO}. In SAPP, the 
trajectories for all cameras are generated as soon as the proposed RML-based distributed path planning algorithm converges at an equilibrium solution, resulting in a per-step execution time that is longer than that of~{}{SMA-NBO}. It is noteworthy that the execution time increases when we plan trajectories over a longer time horizon, i.e., $5$ time steps. Nevertheless {}{the running time of SAPP is not too long, e.g., $1.170$ s when we plan paths for $30$ cameras over a $1$-time-step horizon to observe $50$ targets. It demonstrates that SAPP is not computationally expensive, and it can be deployed in real scenarios.}} 

\begin{table}[h]
\centering
\caption{{}{Results of Wilcoxon rank-sum test (p-value) and Cohen's d effect size for all scenarios. 
The former compares two groups of data without assuming any specific pattern in the data; a p-value of $0.01$ or less is considered as strong evidence for a difference. 
The latter characterises the practical benefits of SAPP over SMA-NBO: an absolute value $d = 0.2$ is considered a ``small'' benefit, $0.5$ represents a ``medium'' benefit, and $0.8$ a large benefit.}}

\renewcommand{\arraystretch}{1.5}
\begin{tabular}
{@{}P{1cm}P{2cm}P{3cm}@{}}\toprule 
  {Scenarios} & {p-value} & {\textit{Cohen's d} effect size}\\
  \midrule
   $1$ & $0.455$ & $-0.03$ \\
   $2$ & $0.636$ & $-0.01$ \\
   $3$ & $0.308$ & $-0.13$ \\
   $4$ & $0.296$ & $-0.13$ \\
   $5$ & $0.003$ & $0.38$ \\
   $6$ & $0.005$ & $0.34$ \\
   $7$ & $<0.001$ & $0.93$\\
   $8$ & $<0.001$ & $0.93$ \\
   $9$ & $<0.001$ & $0.57$\\
   $10$ & $0.008$ & $0.43$ \\
   $11$ & $<0.001$ & $0.86$ \\
   $12$ & $< 0.001$ & $0.91$\\
   \bottomrule
\end{tabular}
\label{table:p_value}\vspace{3mm}
\end{table}

{}{To further characterise the performance differences, we provide in Table~\ref{table:p_value} the outcomes of two statistical tests. As we can see, both approaches perform comparably on the smallest four scenarios. From Scenario $5$ on forward (i.e. from just 5 targets on), SAPP outperforms SMA-NBO with medium to large benefits in our favour.}

\section{Conclusion} \label{sec:conclusion}

In this paper, we have shown a sensor allocation and path planning approach to enhance the awareness of multiple targets in maritime situations. Here, the number of available sensors is lower than the number of targets while their FOV and detection range are limited. An optimization problem is formulated to distribute multiple sensors and plan paths for these sensors over a time horizon. Due to the complexity and large search space of this problem, we cope with it by developing two separate algorithms: a sensor allocation algorithm and a distributed path planning algorithm. The former relies on the sensor-target distance and the duration that targets have not been observed to allocate the sensors. The latter allows each camera to individually plan its trajectory to observe the targets. 

Our experimental results show that compared to the existing solution, SAPP is efficient to minimize the duration to achieve situational awareness improvement under several different scenarios. 

\section*{Acknowledgement}
This project is supported by the Australian Research Council (ARC) project number LP200200881, Safran Electronics and Defense Australasia, and Safran Electronics and Defense.

\balance

\bibliographystyle{IEEEtran}

\bibliography{reference}

\end{document}